\documentstyle[prl,preprint,aps,epsfig]{revtex}
\tightenlines
\begin{document}

\newcommand{\adag}{a^{\dag}}
\newcommand{\atil}{\tilde{a}}
\def\frp#1{${#1\over2}^+$}
\def\frm#1{${#1\over2}^-$}
\def\g{\noindent}

\def\mev{\hbox{\ MeV}}
\def\kev{\hbox{\ keV}}
\def\lambdabar{{\mathchar'26\mkern-9mu\lambda}}
\def\lambdabarrr{{^-\mkern-12mu\lambda}}

\draft
\title{Analysis of the $^{16}O(p,\gamma )^{17}F$
capture reaction using the shell model embedded in the continuum}

\author{K. Bennaceur\dag, F. Nowacki\dag\ddag,  J. Oko{\l}owicz\dag\S ~and
M. P{\l}oszajczak\dag}
\address{\dag\ Grand Acc\'{e}l\'{e}rateur National d'Ions Lourds (GANIL),
CEA/DSM -- CNRS/IN2P3, BP 5027, F-14076 Caen Cedex 05, France}
\address{\ddag\ Laboratoire de Physique Th\'{e}orique  Strasbourg (EP 106),
3-5 rue de l'Universite, F-67084 Strasbourg Cedex, France }
\address{\S\ Institute of Nuclear Physics, Radzikowskiego 152,
PL - 31342 Krakow, Poland}


\maketitle

\begin{abstract}
\parbox{14cm}{\rm We apply the realistic shell model which
includes the coupling between many-particle
(quasi-)bound states and the continuum of one-particle scattering states,
called the Shell Model Embedded in the Continuum, to
the spectroscopy of mirror nuclei: $^{17}$F and $^{17}$O, as well as
to the description of low energy
cross section (the astrophysical $S$ factor) for $E1$, $E2$ and $M1$ components
in the capture reaction:
$^{16}\mbox{O}(p,\gamma)^{17}\mbox{F}$. With the same microscopic input we
calculate the phase shifts and differential cross-section for the elastic
scattering of protons on $^{16}\mbox{O}$.}

\end{abstract}

\vfill
\newpage

\section{Introduction}
Microscopic description of weakly bound exotic
nuclei close to the drip-lines such as, {\it e.g.}, $^8\mbox{B}$ or
$^{11}\mbox{Li}$ in their ground state (g.s.), or nuclei close to the
$\beta$-stability line in the excited configurations like, {\it e.g.},
$^{17}\mbox{F}$ in the first excited state $J^{\pi}=1/2^{+}$,
is an exciting theoretical challenge due to the proximity of particle
continuum. Closeness of the scattering continuum implies that
virtual excitations to continuum states cannot be neglected as they
modify the effective interactions and cause the large spatial dimension
of nuclear density distribution and, in particular, the existence of
halo structures. Even basic concepts of the nuclear collective model hardly
apply for those weakly bound configurations as the particle motion in the
weakly bound orbit is presumably decoupled from the rest of the system
\cite{witek}. The microscopic description of such exotic configurations
should treat properly the residual coupling of discrete configurations
and the scattering continuum. Recently, it was proposed to approach this
difficult problem in the quantum open system formalism which does not separate
the subspaces of (quasi-) bound (the $Q$~subspace) and scattering (the
$P$~subspace) states \cite{bnop1,bnop2}.
Such a formalism can be provided by the Continuum Shell Model (CSM)
\cite{bartz1}
which in the restricted space of configurations generated using the
finite-depth potential has been studied for the giant resonances and for
the radiative capture reactions probing the microscopic structure of
these resonances \cite{bartz1,bartz2,bartz3}.

In this context, capture reaction $^{16}\mbox{O}(p,\gamma )^{17}\mbox{F}$
is interesting for
several reasons. First of all, precise experimental data in the energy
range from 200~keV to 3750~keV are now available \cite{morlock} and
the decays to the g.s.\ ($J^{\pi}=5/2^{+}$) and to the first excited state
($J^{\pi}=1/2^{+}$) of $^{17}\mbox{F}$  have been accurately
resolved \cite{morlock}. The strikingly different behavior of the
astrophysical $S$-factor for the proton capture into the
$5/2^{+}$ state and into the $1/2^{+}$ weakly bound state ($Q=105$ keV) has been
explained by the existence of proton halo in the $1/2^{+}$ state of
$^{17}\mbox{F}$ \cite{morlock}. Different theoretical approaches,
including the potential model \cite{morlock,rolfs},
the model based on the  Generator Coordinate Method
(GCM) \cite{bdh98}, or the $K$- and $R$-matrix analysis \cite{brune}, have been
tried to describe this reaction. Moreover, an expected
simplicity of the low energy wave functions in
$^{16}\mbox{O}$, $^{17}\mbox{O}$ and  $^{17}\mbox{F}$, allows to test
certain salient features of models such as the
effective interactions between $Q$ and $P$ subspaces and the possible quenching
of matrix elements of the residual coupling which depends on
the model space used in the calculations.

Secondly, the exact knowledge of rate for the
reaction $^{16}\mbox{O}(p,\gamma )^{17}\mbox{F}$ is necessary
for modelling of nucleosynthesis process in the hydrogen-burning stars.
Explosive hydrogen burning is believed to occur at various
sites in the Universe, including novas, X-ray bursts, or the
supermassive stars \cite{explo,flowback}.
Hydrogen burning of second-generation stars proceeds mainly through proton -
proton chain and CNO cycle. The changeover from the $pp$ chain to the
CNO cycle happens near $T \simeq 2\cdot
10^{7}$K.  The reaction $^{16}\mbox{O}(p,\gamma )^{17}\mbox{F}$ is of
particular interest in this context as it provides a link to the higher
branches of the CNO cycle. In particular, it starts
the sidebranch : $^{16}\mbox{O}(p,\gamma )^{17}\mbox{F}(p,\gamma
)^{18}\mbox{Ne}({\beta}^{+} \nu)^{18}\mbox{F}(p,\alpha )^{15}\mbox{O}$.
The contribution of CNO cycles to the total amount of produced energy in the
Sun is small and CNO neutrinos account only for about 0.02 of the total neutrino
flux \cite{adel98}. Moreover, most of them are coming from the decay of
$^{13}\mbox{N}$ and $^{15}\mbox{O}$ in the main CNO cycle CNO-I.
Flux of '$^{17}\mbox{F}$-neutrinos', which is again two orders of
magnitude smaller than the flux of neutrinos
from  $^{13}\mbox{N}$ and $^{15}\mbox{O}$
in CNO-I reactions, is controlled by the reaction
$^{16}\mbox{O}(p,\gamma )^{17}\mbox{F}$ in the CNO-II cycle. Hence, the
measurement of CNO
neutrinos coming from different sources provides an accurate
handle on the thermonuclear reaction process in stars like the Sun
and allows in principle to distinct between different
branches of the CNO cycle.

The Shell Model Embedded in the Continuum (SMEC) model, in which
realistic Shell Model (SM) solutions for (quasi-)bound states are coupled to
the one-particle scattering continuum, is a recent development of the
Continuum Shell Model (CSM) \cite{bartz1,bartz2,bartz3} for
the description of
complicated low energy excitations of weakly bound nuclei.
The SMEC approach is based on the {\it realistic} SM which is used
to generate the $N$-particle wave functions. This deliberate choice
implies that the coupling between SM states and the one-particle
scattering continuum has to be given by the residual
nucleon - nucleon interaction. In SMEC, like in the CSM, the bound
(interior) states together with its environment of asymptotic scattering
channels form a quantum closed system. Using
the projection operator technique, one separates the $P$ subspace of
asymptotic channels from the $Q$ subspace of many-body localized
states which are
build up by the bound single-particle (s.p.) wave functions
and by the s.p.\ resonance wave functions. $P$ subspace  is assumed to
contain $(N-1)$-particle states with nucleons on bound s.p.\ orbits
and one nucleon in the scattering state. Also the s.p.\
resonance wave functions outside of the cutoff radius $R_{cut}$~ are
included in the $P$ subspace. The resonance wave functions for
$r < R_{cut}$~ are included in the $Q$ subspace. The
wave functions in $Q$ and $P$ are then properly renormalized
in order to ensure the orthogonality of wave functions in both subspaces.
The application of the SMEC model  for the
description of structure  for mirror nuclei: $^{8}\mbox{B}$,
$^{8}\mbox{Li}$, and capture cross sections for mirror reactions:
$^7\mbox{Be}(p,\gamma)^8\mbox{B}$, $^7\mbox{Li}(n,\gamma)^8\mbox{Li}$
has been published recently \cite{bnop1,bnop2}.

We aim in SMEC at a microscopic description of low lying, complicated
many-body
bound and resonance states. For that reason, the description of particle
continuum is restricted to the
subset of one-nucleon decay channels. This should be a reasonable approach
for describing the structure of mirror nuclei
$^{17}\mbox{F}$ and $^{17}\mbox{O}$ around $^{16}\mbox{O}$,
and the capture reactions:
$^{16}\mbox{O}(p,\gamma)^{17}\mbox{F}$ and
$^{16}\mbox{O}(n,\gamma)^{17}\mbox{O}$. One expects that at higher excitation
energies, {\it e.g.}, above $\alpha$ emission threshold,
the one-particle continuum approximation is too restrictive
and the residual correlations generated in bound state wave functions
by the coupling to those channels cannot be described. Effects of such
correlations have been seen in the structure of $J^{\pi}=3^{+}$ resonances in
$^{8}\mbox{B}$ and $^{8}\mbox{Li}$ \cite{bnop2} which strongly couples to the
three-particle decay channels. In this case, one should try to employ
methods based on the cluster expansion of the wave function and the three-body
continuum models \cite{desc00,csoto,grigorenko}. More complicated
two-particle channels like, {\it e.g.}, the $\alpha$ - decay channel, can be
treated in SMEC, following the approach of Balashov et al. \cite{balashov}.

The paper is organized as follows.  In Sect.~II,
we remind certain elements of the SMEC formalism and, in particular, those
features of the $S$-matrix in SMEC which are involved in the calculation of the
elastic cross-sections, phase shifts and the capture cross-sections.
Sect.~III is devoted to the discussion of specific properties of
$^{17}\mbox{F}$ and $^{17}\mbox{O}$. In particular,
properties of the matrix elements of the effective operator which couple
$Q$ and $P$ subspaces in $^{17}\mbox{F}$ and $^{17}\mbox{O}$ will be discussed
in Sect.~III.B. Features of the
self-consistent average potentials in $Q$ subspace
for different many-body states of $^{17}\mbox{F}$ will be presented in
Sect.~III.C.
Sect.~IV is devoted to the discussion of
$^{16}\mbox{O}(p,\gamma)^{17}\mbox{F}$ capture cross sections for different
multipolarities and different final states of $^{17}\mbox{F}$. The differential
elastic cross-section and the elastic phase-shifts in
$\mbox{p}+{^{16}}\mbox{O}$ scattering will be
compared to the experimental data in Sect.~V. Summary and outlook will be given
in Sect.~VI.

\section{The formalism of Shell Model Embedded in the Continuum}
The full solution of SMEC approach is constructed in three steps. In the first
step, one calculates the (quasi-) bound many-body states in $Q$ subspace. For
that one solves the multiconfigurational SM problem :
\begin{eqnarray}
\label{eq1}
H_{QQ}{\Phi}_i = E_i{\Phi}_i
\end{eqnarray}
where $H_{QQ} \equiv QHQ$~ is
the SM effective Hamiltonian which is appropriate for the SM configuration
space used. For solving (\ref{eq1}), we use the code ANTOINE \cite{caurier}
which employs the Lanczos algorithm and allows for the diagonalization
in large model spaces.

For the coupling between bound and scattering
states around $^{16}\mbox{O}$, we use either a combination of
Wigner and Bartlett forces (WB force) \cite{bnop2} :
\begin{eqnarray}
\label{force}
V_{12} = -V_{12}^{(0)}
[\alpha + (1-\alpha )P_{12}^{\sigma}]\delta({\bf r}_1 - {\bf r}_2) ~ \ ,
\end{eqnarray}
or the density dependent interaction (DDSM1) :
\begin{eqnarray}
\label{force1}
V_{12} = \left[ {\rho}(r){\hat v}_{00}^{in}
(1-{\rho}(r)){\hat v}_{00}^{ex}
{\tau}_1{\cdot}{\tau}_2\left( {\rho}(r){\hat v}_{01}^{in} +
(1-{\rho}(r) ) {\hat v}_{01}^{ex}\right) \right]
\delta({\bf r}_1 - {\bf r}_2)  ~ \ .
\end{eqnarray}
$P_{12}^{\sigma}$ in (\ref{force}) is the spin exchange operator and
${\rho}(r)$ in (\ref{force1}) is  :
\begin{eqnarray}
\label{pot01}
{\rho}(r) = \left[ 1 + \exp \left( \frac{r-r_0}{d}
\right) \right]^{-1} ~ \ .
\end{eqnarray}
with $r_0 = 2.64$ fm and $d = 0.58$ fm. The DDSM1 force
(\ref{force1}) has four parameters determined as: ${\hat v}_{00}^{in} = 36.45$,
${\hat v}_{00}^{ex} = -297.9$, ${\hat v}_{01}^{in} = 109.4$ and
${\hat v}_{01}^{ex} = 115.5$, all in units MeV$\cdot$fm$^{3}$. Similar force
has been used by Schwesinger and Wambach \cite{schw84} for the description of
giant resonances in (1$p$ - 1$h$) - (2$p$ - 2$h$) space. 
However, as compared to their
interaction which is close to the Landau - Migdal type of interactions,
 the DDSM1 parameters ${\hat v}_{00}^{in}$,
${\hat v}_{00}^{ex}$, ${\hat v}_{01}^{in} $ and
${\hat v}_{01}^{ex}$, are reduced by a
constant factor 0.68. Also the parameter $r_0 $ has been
somewhat reduced to better fit the experimental matter radius in oxygen.
For the WB residual interaction
(\ref{force}), the overall strength $V_{12}^{(0)}$  is
adjusted to reproduce as good as possible the energy spectra and the
decay widths of states in nuclei around $^{16}\mbox{O}$. A reasonable
compromise is provided by
$V_{12}^{(0)} = 300$ MeV$\cdot$fm$^{3}$.  The relative contribution of spin
exchange term has been discussed for the
$p$-shell nuclei $^{8}\mbox{B}$ and $^{8}\mbox{Li}$ \cite{bnop2}. It was argued
that the spin exchange part is very small ($(1 - \alpha )
\simeq 0$) for most $p$-shell nuclei but is
expected to increase for heavier nuclei. In present studies, we
use $(1 - \alpha ) = 0.27$, which is a standard value
in the $1s0d$ mass region \cite{bartz1}.
It should be stressed also that matrix elements of $H_{QQ}$ are not modified
by the residual coupling between $P$ and $Q$ because they are
fitted to experimental discrete levels and  narrow
resonances in the continuum and, therefore, they are believed to contain
already those coupling effects.

The SM wave function has an
incorrect asymptotic radial behavior for unbound states.
Therefore, to generate both the radial s.p.\ wave functions in the $Q$ subspace
and the scattering wave functions in $P$ subspace
we use the average potential of SW type
with the spin-orbit part included:
\begin{eqnarray}
\label{pot}
U(r) = V_0f(r) + V_{SO} {\lambdabar}_{\pi}^2 (2{\bf l}\cdot{\bf s})
\frac{1}{r}\frac{d{f}(r)}{dr}  + V_C  ~ \ ,
\end{eqnarray}
where ${\lambdabar}_{\pi}^2 = 2\,$fm$^2$ is the pion Compton wavelength and
$f(r)$ is the spherically symmetrical SW formfactor :
\begin{eqnarray}
\label{pot1}
f(r) = \left[ 1 + \exp \left( \frac{r-R_0}{a} \right) \right]^{-1} ~ \ .
\end{eqnarray}
The Coulomb potential $V_C$ in (\ref{pot})
is calculated for the uniformly charged sphere with radius $R_0$.

For the continuum part, one solves the coupled channel equations:
\begin{eqnarray}
\label{esp}
(E^{(+)} - H_{PP}){\xi}_{E}^{c(+)} \equiv
\sum_{c^{'}}^{}(E^{(+)} - H_{cc^{'}}) {\xi}_E^{c^{'}(+)} = 0 ~  \ ,
\end{eqnarray}
where index $c$~ denotes different channels and $H_{PP} \equiv PHP$~.
The superscript $(+)$ means that boundary
conditions  for incoming wave in the channel $c$ and
outgoing scattering waves in all channels are used.
The channel states are defined by coupling of one
nucleon in the scattering continuum to the many-body SM state in
$(N - 1)$-nucleus. The channel - channel coupling potential in
(\ref{esp}) is :
\begin{eqnarray}
\label{esp2}
H_{cc^{'}} = (T + U ){\delta}_{cc^{'}} + {\upsilon }_{cc^{'}}^{J} ~ \ ,
\end{eqnarray}
where $T$ is the kinetic-energy operator and
$ {\upsilon }_{cc^{'}}^{J}$ is the channel-channel coupling generated by the
residual interaction. The potential for channel $c$
in (\ref{esp2}) consists of 'initial guess', $U(r)$, and diagonal
part of coupling potential ${\upsilon }_{cc}^{J}$
which depends on both the s.p.\ orbit
${\phi}_{l,j}$ and the considered many-body
state $J^{\pi}$. This modification of the initial potential $U(r)$ change the
generated s.p.\ wave functions ${\phi}_{l,j}$ defining $Q$ subspace which in
turn modify the diagonal part of the residual force, {\it etc.}
This means that the solution of coupled channel equations (\ref{esp}) is
accompanied by the self-consistent iterative procedure which yields for each
channel independently the new {\it self-consistent potential} :
\begin{eqnarray}
\label{usc}
U^{(sc)}(r) = U(r)+{\upsilon }_{cc}^{J(sc)}(r) ~ \ ,
\end{eqnarray}
and consistent with it the
new renormalized matrix elements of the
coupling force. The parameters of the {\it initial potential} $U(r)$
are chosen  in such a way that $U^{(sc)}(r)$
reproduces energies of experimental s.p.\ states,
whenever their identification is possible.

The third system of equations in SMEC consists of inhomogeneous
coupled channel equations:
\begin{eqnarray}
\label{coup}
(E^{(+)} - H_{PP}){\omega}_{i}^{(+)} = H_{PQ}{\Phi}_i \equiv w_i
\end{eqnarray}
with the source term $w_i$ which is primarily given by the
structure of $N$ - particle SM wave function ${\Phi}_i$~. The explicit
form of this source is given in \cite{bnop2}.
These equations define functions ${\omega}_{i}^{(+)}$~,  which
describe the decay of quasi-bound state ${\Phi}_i$~ in the continuum.
The source $w_i$~ couples the wave function of $N$-nucleon
localized states with $(N-1)$-nucleon localized states plus one nucleon
in the continuum. Formfactor of the source term is given by
the self-consistently determined s.p.\ wave functions.

The full solution of SMEC equations is expressed by three functions:
${\Phi}_i$~, ${\xi}_{E}^{c}$~ and ${\omega}_i$ \cite{bnop2,bartz1}~:
\begin{eqnarray}
\label{eq2}
{\Psi}_{E}^{c} = {\xi}_{E}^{c} + \sum_{i,j}({\Phi}_i + {\omega}_i)
\frac{1}{E - H_{QQ}^{eff}}
<{\Phi}_{j}\mid H_{QP} \mid{\xi}_{E}^{c}> ~ \ ,
\end{eqnarray}
where
\begin{eqnarray}
\label{eq2a}
H_{QQ}^{eff} = H_{QQ} + H_{QP}G_{P}^{(+)}H_{PQ} ~ \ ,
\end{eqnarray}
is the {\it effective}
SM Hamiltonian which includes the coupling to the continuum, and
$G_{P}^{(+)}$~ is the Green function for the motion of s.p.\ in
the $P$ subspace. Matrix $H_{QQ}^{eff}$ is non-Hermitian
(the complex, symmetric matrix)  for energies above the particle
emission  threshold and Hermitian (real) for lower energies.
The eigenvalues ${\tilde {E_i}} - \frac{1}{2}i{\tilde {{\Gamma}_i} }$ are
complex for decaying states and
depend on the energy $E$ of particle in the continuum.
The energy and width of resonance states are determined by the condition:
$\tilde{E_i}(E) = E$ \cite{bartz1}. The eigenstates
corresponding to these eigenvalues can be obtained by the orthogonal but in
general non-unitary transformation :
\begin{eqnarray}
\label{eq2aa}
 {\tilde \Phi}_i = \sum_{j}^{} \beta_{ij} {\Phi}_j
\end{eqnarray}
where the coefficients $\beta_{ij}$ form a complex matrix of eigenvectors
in the SM basis satisfying :
\begin{eqnarray}
\label{eq2ab}
\sum_k \beta_{ik} \beta_{jk} = \delta_{ij}
\end{eqnarray}
Analogously, one can define :
\begin{eqnarray}
\label{eq2ac}
{\tilde \omega}_i = \sum_j \beta_{ij} \omega_j
\end{eqnarray}
and
\begin{eqnarray}
\label{eq2ad}
{\tilde \Omega}_i = {\tilde \Phi}_i + {\tilde \omega}_i  ~ \ .
\end{eqnarray}
Inserting them in (\ref{eq2}), one obtains :
\begin{eqnarray}
\label{cons}
{\Psi}_{E}^{c} = {\xi}_{E}^{c} + \sum_{i}^{}{\tilde {\Omega}_i}
\frac{1}{E - {\tilde E_i}
+ (i/2){\tilde {\Gamma}_i}} <{\tilde {\Phi}_i} \mid H_{QP} \mid {\xi}_{E}^{c}>
\end{eqnarray}
for the continuum many-body wave function projected on channel $c$, where
\begin{eqnarray}
\label{diss}
{\tilde {\Omega}_i} = {\tilde {\Phi}_i} + \sum_{c}
\int_{{\varepsilon}_c}^{\infty} dE^{'} {\xi}_{E^{'}}^{c}
\frac{1}{E^{(+)} - E^{'}}
<{\xi}_{E^{'}}^{c}\mid H_{PQ} \mid {\tilde {\Phi}_i}> ~ \ ,
\end{eqnarray}
is  the wave function of discrete state modified by the coupling to the
continuum states. It should be stressed that the
SMEC formalism is {\it fully symmetric} in treating the
continuum and bound state parts of the solution : ${\Psi}_{E}^{c}$
represents the continuum state modified by the discrete states, and
${\tilde {\Omega}_i}$ represents the discrete
state modified by the coupling to the continuum.

\subsection{Radiative capture and electromagnetic transitions}
Having obtained the full solution of SMEC, one can calculate many
observable quantities. {\it E.g.}, the calculation of capture cross-section
goes as follows. The initial SMEC wave function for the
$[\mbox{p} \bigotimes (N-1)]^{J_{i}^{{\pi}_i}}$ system is :
\begin{eqnarray}
\label{psiin}
\Psi_i(r)=\sum_{l_a j_a}i^{l_a}{\psi_{l_a j_a}^{J_i}(r)\over r}
\biggl[\bigl[Y^{l_a}\times\chi^{s}\bigr]^{j_a}\times\chi^{I_t}\biggr]^{(J_i)}
_{m_i} ~~~ \ ,
\end{eqnarray}
and the final SMEC wave function for the $N$-system in
the many-body state $J_{f}^{{\pi}_f}$ is :
\begin{eqnarray}
\label{psifin}
\Psi_f(r)=\sum_{l_b j_b}A_{l_bsj_b}^{j_bI_bJ_f} {u_{l_b j_b}^{J_f}(r)\over r}
\biggl[\bigl[Y^{l_b}\times\chi^{s}\bigr]^{j_b}\times\chi^{I_t}\biggr]^{(J_f)}
_{m_f}  \ .
\end{eqnarray}
$I_t$~ and $s$~ denote the spin of target nucleus and
incoming proton, respectively.
$A_{l_bsj_b}^{j_bI_bJ_f}$~ is the coefficient of fractional parentage and
$u_{l_bj_b}^{J_f}$~ is the s.p.\ wave in the many-particle state~$J_f$~.
These SMEC wave functions (\ref{psiin},\ref{psifin}) are then used to
calculate the transition amplitudes $T^{E{\cal L}}$ and $T^{M1}$ for
$E1$~, $E2$~ and $M1$ transitions, respectively \cite{bnop1,bnop2}.
The radiative capture cross section is :
\begin{eqnarray}
\label{tran}
\sigma^{E1,M1} = {16\pi\over9} \biggl({k_\gamma\over k_p}\biggr)^3
  \biggl({\mu\over\hbar c}\biggr)
  \biggl({e^{2}\over\hbar c}\biggr)
  {1\over 2s+1}~{1\over 2I_t+1}
\sum\mid T^{E1,M1}\mid^2
\end{eqnarray}
\begin{eqnarray}
\label{tran1}
\sigma^{E2} = {4\pi\over75} \biggl({k_\gamma^5\over k_p^3}\biggr)
  \biggl({\mu\over\hbar c}\biggr)
  \biggl({e^{2}\over\hbar c}\biggr)
  {1\over 2s+1}~{1\over 2I_t+1} \sum\mid T^{E2}\mid^2
\end{eqnarray}
and $\mu$~ stands for the reduced mass of the system.

Another interesting quantities are $B(E\lambda )$, $B(M\lambda )$
transition matrix elements between the SMEC wave functions $J_i^{{\pi}_i}$
and $J_f^{{\pi}_f}$ for the [$ N $] - system  :
\begin{eqnarray}
\label{be2}
B(\Pi \lambda ; J_{i}^{\pi_i} \rightarrow J_{f}^{\pi_f}) =
\frac{\mid <\Psi_f(J_{f}^{{\pi}_f}) || {\hat O}(\Pi \lambda ) ||
\Psi_i(J_{i}^{{\pi}_i}) >{\mid}^2}{2J_i + 1} ~~~ \ ,
\end{eqnarray}
where $\Pi\,=\,$ E (electric) or M (magnetic), and
$\lambda\,=\, 1,2...$ (multipolarity). ${\hat O}(\Pi \lambda )$ in (\ref{be2})
is the electromagnetic transition operator \cite{brussard}.

\subsection{Properties of the $S$-matrix }
The $S$-matrix for the scattering of a nucleon by a nucleus is given by the
asymptotic behavior of the total wave function (\ref{cons}) :
\begin{eqnarray}
\label{cons1}
{\Psi}_{c}^{(c_0)} = {\xi}_{c}^{(c_0)} + \sum_{i}^{}{\tilde {\Omega}_i}
\frac{1}{E - {\tilde E_i}
+ (i/2){\tilde {\Gamma}_i}} <{\tilde {\Phi}_i} \mid H \mid {\xi}_{c}^{(c_0)}>
\end{eqnarray}
with the incoming wave only in the channel $c_0$. The asymptotic conditions for
the solution (\ref{cons1}) have been analyzed in Ref. \cite{bartz1}. Here we
only give the final result for the amplitude of the partial width :
\begin{eqnarray}
\label{par1}
{\gamma}_c^{(n)}=-\exp (-i{\delta}_c^{(0)})
\left( \frac{4m_r}{{\hbar}^2k_c} \right)^{1/2} \sum_{i}^{}{\beta}_{ni}
\sum_{c^{'}}^{} \int dr
{\xi}_{c^{'}}^{(c)}(r)w_{c^{'}}^{(i)}(r) ~~~ \ ,
\end{eqnarray}
where ${\delta}_c^{(0)}$ is the background scattering phase :
\begin{eqnarray}
\label{par2}
 {\delta}_c^{(0)} = \frac{1}{2} \mbox{arg}S_{cc}^{(0)} ~ \ ,
\end{eqnarray}
and $S_{cc}^{(0)}$ is determined from the asymptotic, large distance
behavior of
${\xi}_{c}^{(c)}$. Using the proportionality relation between matrix elements
in (\ref{cons1}) and the amplitudes  of partial width  (\ref{par1}),
one derives the $S$-matrix elements :
\begin{eqnarray}
\label{par3}
S_{cc_{0}} = S_{cc_{0}}^{(0)} -i\exp [i({\delta}_{c}^{(0)} +
{\delta}_{c_{0}}^{(0)})]
\sum_{n} \frac{{\gamma}_{c}^{(n)}{\gamma}_{c_{0}}^{(n)}}{E - {\tilde E}_n +
\frac{1}{2}i{\tilde {\Gamma}}_n} ~~~ \ .
\end{eqnarray}
Amplitudes of the partial widths ${\gamma}_{c}^{(n)}$ as well as the real
${\tilde E}_n$ and imaginary ${\tilde {\Gamma}}_n$ parts of complex
eigenvalues of the effective shell model Hamiltonian $H_{QQ}^{eff}$
(\ref{eq2a}) entering the
$S$-matrix elements are explicitly energy dependent. ${\gamma}_{c}^{(n)}$ are
also complex due to the channel - channel coupling and the mixing of
quasi-bound states embedded in the continuum \cite{bartz1}.
The partial width for channel $c$ can then be defined by :
\begin{eqnarray}
\label{par4}
{\Gamma}_{n,c} = {\mid}{\gamma}_{c}^{(n)}{\mid}^2 ~~~ \ ,
\end{eqnarray}
though the total width ${\Gamma}_{n}$ is no longer sum of partial widths
\cite{mahaux} :
\begin{eqnarray}
\label{par5}
\sum_{c}^{}{\mid}{\gamma}_{c}^{(n)}{\mid}^2 = {\tilde {\Gamma}}_{n} \sum_{i}
{\mid}{\beta}_{ni}{\mid}^2 \geq {\tilde {\Gamma}}_{n}
\end{eqnarray}
because the eigenvectors $\beta_{ni}$ are normalized in a sense of
(\ref{eq2ab}), what implies :
\begin{eqnarray}
\label{par5a}
\sum_i \mid \beta_{ni} {\mid}^2 \geq 1  ~ \ .
\end{eqnarray}

\section{Spectroscopy of $^{17}\mbox{F}$ and $^{17}\mbox{O}$ nuclei}

\subsection{The effective interactions for the $0p1s1d$ SM space}
For the purpose of the present study we have constructed SM effective
interactions in the cross-shell model space connecting the $0p$ and $1s0d$
shells. The interactions have three distinctive parts: the $0p$-shell part
taken to be the CK(8-16) interaction \cite{cohen},
the $1s0d$-shell part taken to be the Brown--Wildenthal interaction
\cite{brownwil}. For the cross-shell matrix elements, we use the G matrix
of Kahana, Lee and Scott (KLS) \cite {kls}.
This part of the interaction has to be modified phenomenologically.
We proceed as follow: the $1s0d$ s.p.\ energies above the $^4$He
core are taken from  generalized monopoles fit  
of  energies of single-particle and single-hole states over the mass table 
\cite{gemo}.
Then the cross monopoles are adjusted to reproduce $^{15}$N, $^{15}$O and
$^{17}$O, $^{17}$F spectra.
The center of mass components are removed with the usual Gloeckner - Lawson
prescription \cite{lawson} adding a center of mass kinetic energy hamiltonian.
For natural parity states, we consider only 0$\hbar \omega$
excitations and for non natural parity states, 1$\hbar \omega$
excitations.
Of course, it is well known and has been studied in the past, that
$^{16}$O (and$^{17}$O, $^{17}$F) have significant amount of $2p - 2h$
($3p - 2h$) and $4p - 4h$ ($5p - 4h$) admixtures in their low lying states
\cite{brown,zbm}. Nevertheless, the shell model calculation of these admixtures
is not straightforward:
\begin{itemize}
\item Firstly, it has been already stressed in the past by different
  authors \cite{zamick,sdpf,wbm}, that $N \hbar \omega$ mixing converges
slowly with $N$ and, if stopped for example after $N=2$,
causes a strong lowering of $0p - 0h$ states due to 2$\hbar \omega$
pairing correlations which were already present in pure $2p - 2h$ states.
One could overpass this artifact of the calculation by artificially
lowering the $N \hbar \omega$ unperturbed configurations through the
monopoles of the interaction, but the control of convergence of
the mixing would be impossible as higher order admixtures would
destroy the picture.
\item Secondly, the building blocks of our interaction are made of
0$\hbar \omega$ $0p$ and $1s0d$ phenomenological interactions,
which in principle already contain, in particular through pairing
renormalizations, the higher order correlations.
For this reason, we have decided to fix the unperturbed ($4p - 4h$) 0$^+$ state
around its experimental energy (6.049 MeV).
\end{itemize}

Again,  the derivation of precise correlated wave functions for
$^{16}$O and $^{17}$O, $^{17}$F is out of the scope of this paper and we will
refer to the mixing derived by other studies \cite{brown,zbm}.

\subsection{The effective operator of the residual coupling}
Expressions for the matrix elements of residual interaction which couples
$P$ and $Q$ subspaces have been given in Refs. \cite{bartz1,bnop2}. There are
two kinds of coupling operators. The channel - channel coupling reduced
matrix elements involve one-body operators :
\begin{eqnarray}
\label{ef1}
{\cal O}^{K}_{{\beta}{\delta}} =
(a^{\dagger}_{\beta}{\tilde a}_{\delta})^K
\end{eqnarray}
Diagonal part of ${\cal O}$ induce the
renormalization of the s.p.\ average potential.
Matrix elements of the operators ${\cal O}$  are calculated between
different many-body states in the $(N - 1)$-system,  {\it i.e.}, in
$^{16}\mbox{O}$, and they depend sensibly on the amount of $2p - 2h$ and
$4p - 4h$ correlations in the g.s.\ of $^{16}\mbox{O}$.

Reduced matrix elements of the source term
$w_i$ in the inhomogeneous coupled channel equations (\ref{coup})
contain a product of two
annihilation operators and one creation operator :
\begin{eqnarray}
\label{ef2}
{\cal R}^{j_{\alpha}}_{{\gamma}{\delta}(L){\beta}}=
(a^\dagger_{\beta}({\tilde a}_{\gamma}{\tilde a}_{\delta})^L)^{j_{\alpha}}
      ~~~~ \ .
\end{eqnarray}
Operators ${\cal R}$ enter in the calculation of
complex eigenvalues of $H_{QQ}^{eff}$.
Matrix elements of ${\cal R}$ are calculated between
different initial state wave functions in either
[$\mbox{p} \bigotimes (N-1)$] or
[$\mbox{n} \bigotimes (N-1)$] system, {\it i.e.}, in $^{17}\mbox{F}$ or
$^{17}\mbox{O}$, and a given final state wave
function in $^{16}\mbox{O}$.
These matrix elements depend on the configuration mixing in the many-body
wave functions of $^{17}\mbox{F}$, $^{17}\mbox{O}$ and $^{16}\mbox{O}$.

We perform the SM calculations for $^{16}\mbox{O}$ ground state in which
admixture of $2p - 2h$ and $4p - 4h$ configurations are neglected :
\begin{eqnarray}
\label{ef3}
|^{16}\mbox{O}, 0_1^+ > & = & a_{00}{|(0p - 0h)^{J=0} >} +
a_{22}{|(2p - 2h)^{J=0} >} +
a_{44}{|(4p - 4h)^{J=0} >} +  \cdots \nonumber \\
& \simeq & a_{00}{|(0p - 0h)^{J=0} >}
\end{eqnarray}
{\it i.e.},  the g.s.\ of $^{16}\mbox{O}$
is taken to be a pure $(0p^{12})$ $0p - 0h$ configuration and the second excited
$0_2^{+}$ state with predominantly $4p - 4h$ structure is not mixed with the
ground state.
How big are these $\mbox{n}p - \mbox{n}h$ admixture depends on the
SM effective interaction and the effective s.p.\ space
used \cite{brown,zbm}. In Table~\ref{decompo1} we give
results of different SM calculations for $0_1^{+}$ in  $^{16}\mbox{O}$.

Likewise in $^{17}\mbox{O}$ and $^{17}\mbox{F}$, the
g.s.\ $J^{\pi}=5/2_{1}^{+}$ and the excited states $J^{\pi}=1/2_{1}^{+}$,
$3/2_{1}^{+}$ correspond in our approximation
to the pure configuration of one particle in, respectively,
$0d_{5/2}$, $1s_{1/2}$ or $0d_{3/2}$ shells outside of the $(0p^{12})$-core of
$^{16}\mbox{O}$, {\it i.e.\/}:
\begin{eqnarray}
\label{ef4}
|^{17}\mbox{F}; J^{\pi}, {\pi = +} > & = &
a_{10}^{J^{\pi}} {|(1p - 0h)^{J^{\pi}} >}
+ a_{32}^{J^{\pi}} {|(3p - 2h)^{J^{\pi}} >} + \cdots \nonumber \\
& \simeq & a_{10}^{J^{\pi}} {|(1p - 0h)^{J^{\pi}} >} ~ \ .
\end{eqnarray}
The amount of these admixture calculated by Brown--Green \cite{brown} and
Zuker--Buck--McGrory \cite{zbm} can be read from the
Table~\ref{decompo2}. The experimental spectroscopic amplitudes for $5/2_1^{+}$
and $1/2_1^{+}$ states can be found in Table~\ref{decompo3}.

The negative parity states involve the excitations across the shell
closure and involve one hole in the $p$ shell and two particles in the
$1s0d$ shells. In our model of $^{16}\mbox{O}$ where $2p - 2h$ and $4p - 4h$
configurations are neglected, the spectroscopic factors of
negative parity SM states are equal zero.
Hence the particle decay width for these states is coming
from the functions $\omega_i$  (\ref{coup})
which describe the continuation of $Q$ space wave functions $\Phi_i$ into $P$,
{\it i.e.}, from the modification of discrete states by the coupling to the
continuum (\ref{diss}).

Neglecting higher order correlations in the g.s.\ of $^{16}\mbox{O}$ and in
positive parity excited states of either
$^{17}\mbox{F}$ or $^{17}\mbox{O}$ means that
residual coupling between $Q$ and $P$ which involves those states
should be properly rescaled, {\it i.e.},
the matrix elements of $\cal{O}$, $\cal{R}$ should be quenched  :
\begin{eqnarray}
\label{ef4a}
<{^{16}\mbox{O}}, J_f| {\cal O} |{^{16}\mbox{O}}, J_i>^{(quenched)} =
(a_{00})^2 <{^{16}\mbox{O}}, J_f| {\cal O} |{^{16}\mbox{O}}, J_i>^{(non-quenched)}
\end{eqnarray}
\begin{eqnarray}
\label{ef4b}
<{^{16}\mbox{O}}, J_f| {\cal R} |{^{17}\mbox{O}}, J_i>^{(quenched)} =
a_{00}a_{10}^{J_i} <{^{16}\mbox{O}}, J_f| {\cal R} |{^{17}\mbox{O}}, J_i>^{(non-quenched)}
\end{eqnarray}
\begin{eqnarray}
\label{ef4c}
<{^{16}\mbox{O}}, J_f| {\cal R} |{^{17}\mbox{F}}, J_i>^{(quenched)} =
a_{00}a_{10}^{J_i} <{^{16}\mbox{O}}, J_f| {\cal R} |{^{17}\mbox{F}}, J_i>^{(non-quenched)}
\end{eqnarray}
Amount of configuration mixing for negative parity states is large (two
particles in $1s0d$ shell and 1 hole in $0p$ shell) and therefore matrix
elements of ${\cal R}$ are not quenched.
In our calculations, we take the values given by
Brown - Green \cite{brown} for the amplitudes $a_{00}$  and $a_{10}^{J_i}$
(see Tables~\ref{decompo1} and \ref{decompo2}).

\subsection{The self-consistent average potentials}
Construction of the $Q$ subspace in SMEC is achieved by the
self-consistent, iterative  procedure which
for a given initial average s.p.\ potential (\ref{pot})
and for a given residual two-body interaction between $Q$ and $P$
yields the self-consistent s.p.\ potential $U^{(sc)}(r)$ which depends on the
s.p.\ wave function ${\phi}_{l,j}$, the total spin $J$ of the $N$-nucleon
system as well as on the one-body matrix elements of $(N-1)$ - nucleon
daughter system. As explained above, the positive parity states in
$^{17}\mbox{F}$ and $^{17}\mbox{O}$ are described
as a pure one particle configuration in $1s0d$ shells. Consequently,
the spectroscopic factors for the states $J^{\pi}={1/2}^{+}~, {5/2}^{+}~$  and
${3/2}^{+}~$ are equal 1. Consistently with this approximation,
we can identify position of proton (neutron) s.p.\ orbits
$1s_{1/2}$, $0d_{5/2}$ and $0d_{3/2}$ with, respectively,
$J^{\pi}={1/2}^{+}$, ${5/2}^{+}~$  and ${3/2}^{+}~$
many body states of $^{17}\mbox{F}$ ($^{17}\mbox{O}$), {\it i.e.}, we
may ask that $U^{(sc)}(J^{\pi})$  provides  the energy of s.p.\
orbit at the energy of the corresponding many-body state with respect
to the threshold.
This choice is essential for quantitative description of radiative
capture cross-section.  Consequently,
the initial SW potentials $U(J^{\pi}$) have all different depth parameters.
Moreover, they depend on the choice of the residual interaction coupling $P$
and $Q$. On the contrary, we keep a common radius $R_0 = 3.214\,$fm, a common
diffuseness $a = 0.58\,$fm
and a common  spin-orbit strength $V_{SO} = 3.683\,$MeV for all of them.
The depth parameters of initial SW potentials for
[$\mbox{p} \bigotimes {}^{16}\mbox{O}$] and
[$\mbox{n} \bigotimes {}^{16}\mbox{O}$] systems,
are summarized in Table~\ref{parameters}. For a given nucleus
($^{17}\mbox{F}$ or $^{17}\mbox{O}$),
all those different central potentials correspond  to the one and the same
equivalent potential $U^{(eq)}(r)$ in the SW form which binds s.p.\ orbits
at the same energy as obtained in the self-consistent potentials for
neutrons and protons. For both
WB and DDSM1 residual interactions,
the depth of $U^{(eq)}$ is $V_0 = -52.46\,$MeV in $^{17}\mbox{F}$ and
$V_0 = -52.49\,$MeV in $^{17}\mbox{O}$. The potential radius, the
diffuseness and the spin-orbit strength are the same as used for the initial
potentials $U(J^{\pi})$. The spectrum of single-particle (-hole)
energies in this potential is given in Table~\ref{spen} for $^{17}\mbox{F}$ and
in Table~\ref{spen1} for $^{17}\mbox{O}$. These energies are
determined from the data of $^{17}\mbox{F}$, $^{17}\mbox{O}$  (for particles)
and $^{15}\mbox{N}$, $^{15}\mbox{O}$  (for holes). The difference of
s.p.\ energies ${\epsilon}_p(1s_{1/2}) - {\epsilon}_p(0d_{5/2}) = 0.495\,$MeV
in $^{17}\mbox{F}$ and
${\epsilon}_n(1s_{1/2}) - {\epsilon}_n(0d_{5/2}) = 0.871\,$MeV,
which should be compared with the excitation energy of $J^{\pi}=1/2_{1}^+$
state in those nuclei, is a manifestation of the Thomas-Ehrman shift
\cite{thomas} and is included effectively in
our equivalent and self-consistent s.p.\ potentials for all studied channels.

\begin{figure}[t]
\centerline{\epsfig{figure=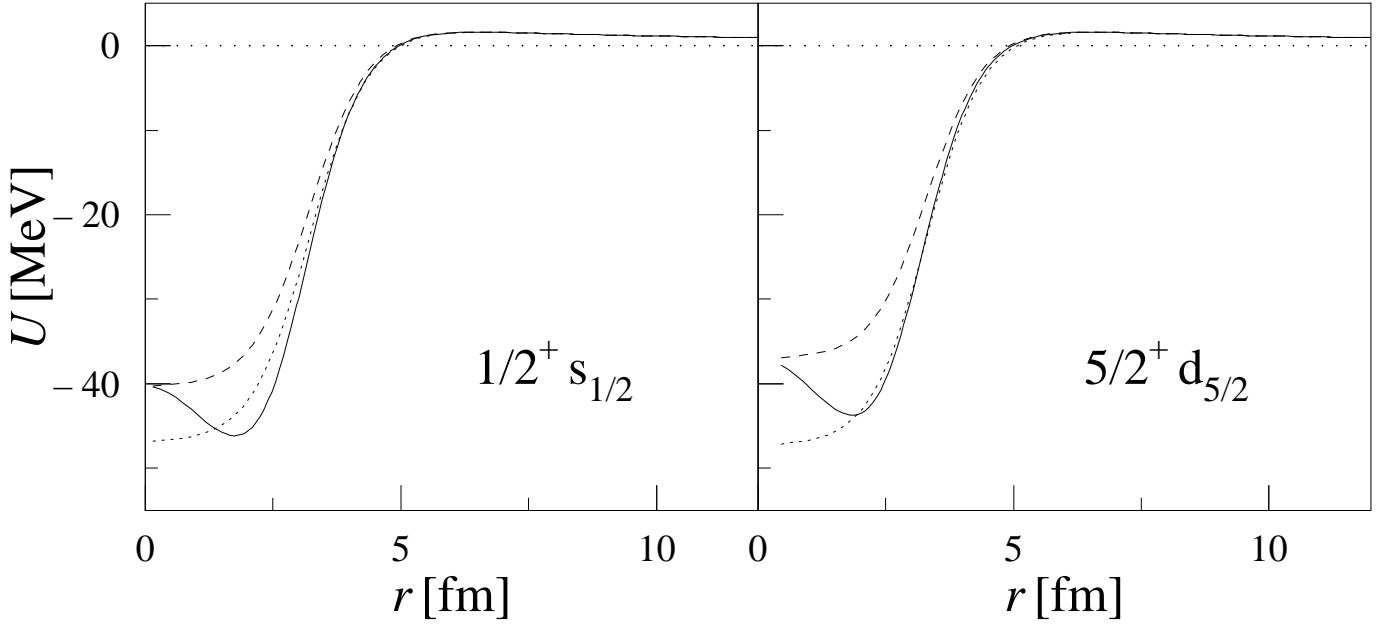,height=9cm}}
\vskip 1truecm
\caption{The self-consistent potentials for $1s_{1/2}$ s.p.\ orbit in
$J^{\pi} = 1/2^{+}$ many body states (the l.h.s.\ picture)
and $0d_{5/2}$ s.p.\ orbit in
$J^{\pi} = 5/2^{+}$ many body states (the r.h.s.\ picture)
in $^{17}\mbox{F}$ are calculated for
the DDSM1 residual interaction (\protect\ref{force1}).}
\label{fig0d}
\end{figure}

In Fig.~\ref{fig0d} we show examples of calculated potentials in
$^{17}\mbox{F}$
for the proton s.p.\ orbitals $1s_{1/2}$ and $0d_{5/2}$
in the total spin states:
$J^{\pi} = 1/2^+$ and $J^{\pi} = 5/2^+$, respectively. The calculations
have been performed using the appropriate initial potentials $U(J^{\pi})$
(see Table~\ref{parameters}) for
the DDSM1 residual interaction (\ref{force1}).
For example, potential $U(5/2^{+})$ is chosen in such a way that the
self-consistent potential $U^{(sc)}(5/2^{+})$ (see Table~\ref{spen}) yields
$0d_{5/2}$ proton s.p.\ orbit bound at the
experimental binding energy  of the ground state (g.s.)
$J^{\pi}={5/2}_{1}^{+}$. Similarly, the choice of $U(1/2^{+})$ and the
determination of $U^{(sc)}(1/2^{+})$ is associated with the reproduction of
the experimental binding energy  of the first excited state
$J^{\pi}={1/2}_{1}^{+}$. The $J^{\pi}$-independent
equivalent potential yields the $0d_{5/2}$ and $1s_{1/2}$ proton s.p.\ orbits
at the position given for them in the corresponding  $U^{(sc)}(5/2^{+})$ and
$U^{(sc)}(1/2^{+})$  potentials.

The self-consistent potential (the solid line) strongly deviates from the SW
form. In the center $U^{(sc)}(r)$ has
a strong maximum  which is absent in the initial potential
$U(r)$ (the dashed line). The self-consistent
potentials $U^{(sc)}(1/2^{+})$ and $U^{(sc)}(5/2^{+})$ are different
(compare the solid curves on l.h.s.\  and
r.h.s.\ of Fig.~\ref{fig0d}), in spite of the fact that up to spin--orbit
coupling the equivalent
potential $U^{(eq)}$ (the dotted lines) in these states
is identical. This clearly shows how strong is the
state dependence of both the self-consistent average fields and
the renormalized matrix elements of the coupling force.
 One should also notice the difference in surface region
between $U^{(sc)}(1/2^{+})$, $U^{(sc)}(5/2^{+})$ and
$U^{(eq)}$. In $U^{(sc)}(5/2^{+})$ one may notice
decrease of the potential radius in comparison with the radius of
$U^{(eq)}$. However,  as compared to the initial average
potentials $U(5/2^{+})$ and $U(1/2^{+})$, both the
self-consistent potentials $U^{(sc)}(5/2^{+})$, $U^{(sc)}(1/2^{+})$, and
the equivalent potential $U^{(eq)}$ are deeper and their effective radii are
bigger.

\begin{figure}[t]
\centerline{\epsfig{figure=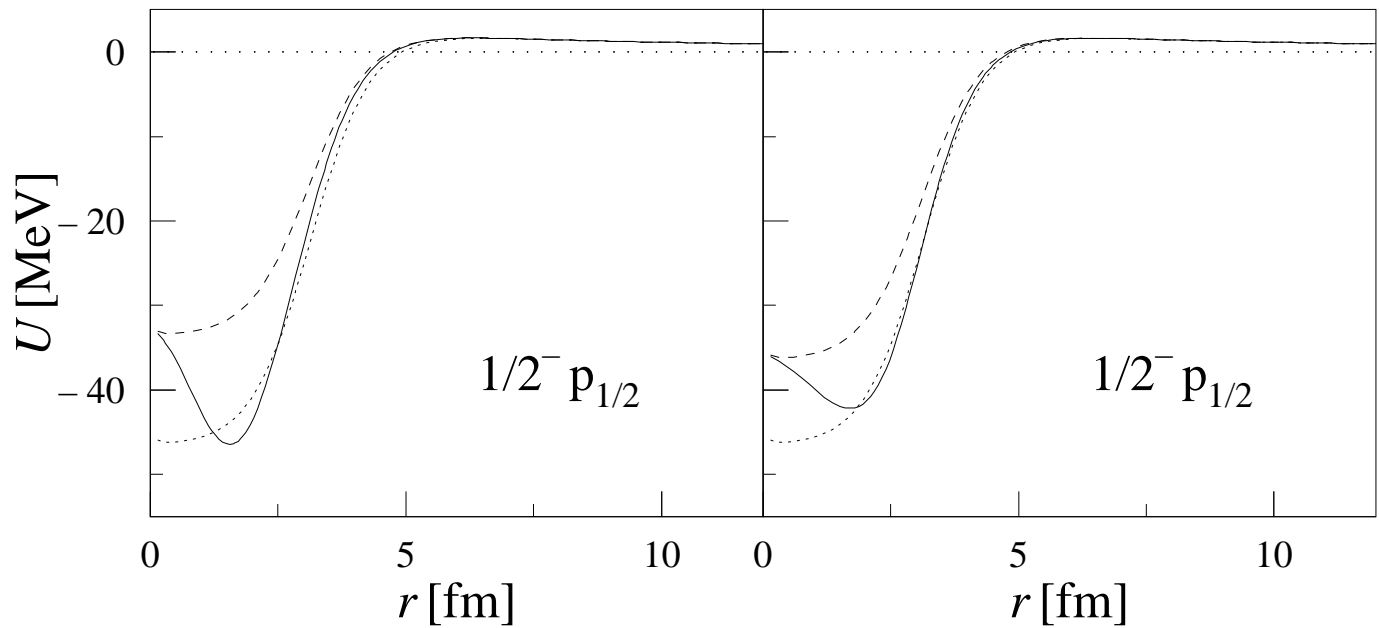,height=9cm}}
\vskip 1truecm
\caption{The self-consistent potentials for $0p_{1/2}$ s.p.\ orbit in
$J^{\pi} = 1/2^{-}$ many body states in $^{17}\mbox{F}$ are calculated with
the WB residual interaction~(\protect\ref{force}) (l.h.s.\ of the plot)
and the DDSM1 interaction~(\protect\ref{force1}) (r.h.s.\ of the plot). }
\label{fig0x}
\end{figure}

In Fig.~\ref{fig0x} we show examples of calculated potentials in
$^{17}\mbox{F}$ for the proton s.p.\ orbital $0p_{1/2}$
in the many-body states  $J^{\pi} = 1/2^-$.
The calculations have been performed using the initial potentials
$U(1/2^{-})$ for
the WB (l.h.s.\ of the plot) and DDSM1 (r.h.s.\ of the plot)
residual interactions. For both interactions, the equivalent
potential $U^{(eq)}(r)$ is the same. It is interesting to notice how different
are the self-consistent $U^{(sc)}(r)$ potentials for the two considered
residual interactions. As a rule, the renormalization of
initial s.p.\ potential is
weaker for the density-dependent DDSM1 interaction. Similarly as for
the self-consistent potentials in $J^{\pi}={5/2}^{+}$ and
$J^{\pi}={1/2}^{+}$ states (see Fig.~\ref{fig0d} ), the effective radius of
$U^{(sc)}(r)$  shrinks as compared to the radius of $U^{(eq)}(r)$
. Again this effect is stronger for the WB residual interaction.

In general, the surface region of average potential shows weak sensitivity
to the self-consistent correction.
In our case, the radial dependence of
self-consistent correction for all positive parity states in $^{17}\mbox{F}$
and $^{17}\mbox{O}$ is  given by the $0p_{1/2}$ and $0p_{3/2}$ radial
formfactors
corresponding to the well bound single-hole state and not by the weakly bound
single-particle states $1s_{1/2}$ or $0d_{5/2}$.
For that reason, even for the $J^{\pi}={1/2}_{1}^{+}$ state in  $^{17}\mbox{F}$
which is bound by 105 keV, induced renormalization of the surface in
the self-consistent potential decreases the radius of potential in the
surface region.

\subsection{Discussion}
In this section, we shall present the SMEC
results for the spectrum of mirror nuclei: $^{17}\mbox{F}$ and
$^{17}\mbox{O}$. The results depend mainly on:
(i) the nucleon - nucleon interaction in $Q$ subspace,
(ii) the residual coupling of
$Q$ and $P$ subspaces, (iii) the self-consistent
average s.p.\ potential which generates the radial
formfactor for s.p.\ bound wave functions and s.p.\ resonances,
and (iv) the cutoff radius for s.p.\ resonances. Below, we shall comment on
their relative importance.

\subsubsection{Spectrum of $^{17}\mbox{F}$}
\begin{figure}[t]
\centerline{\epsfig{figure=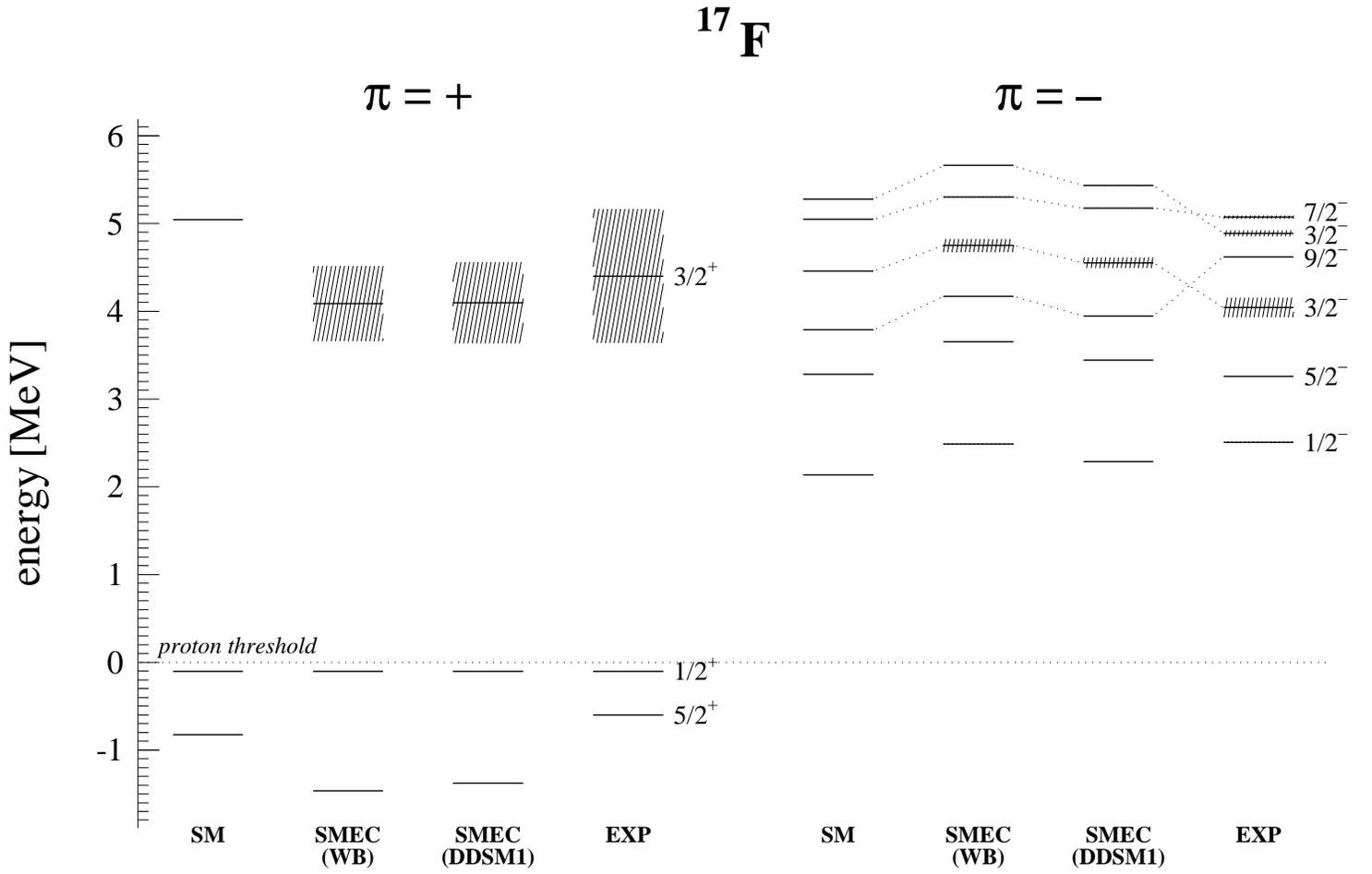,height=14cm}}
\vskip 1truecm
\caption{SM and SMEC in different approximations
labelled 'SMEC (WB)', SMEC (DDSM1)' vs.\ experimental states of
$^{17}\mbox{F}$ nucleus.  The proton threshold
energy is adjusted to reproduce position of the $1/2_{1}^{+}$
first excited
state. The shaded regions represent the width of resonance states. For the
details of the calculation see the description in the text and in the caption
of Table~\protect\ref{f17}.}
\label{figf17}
\end{figure}

\begin{figure}[t]
\centerline{\epsfig{figure=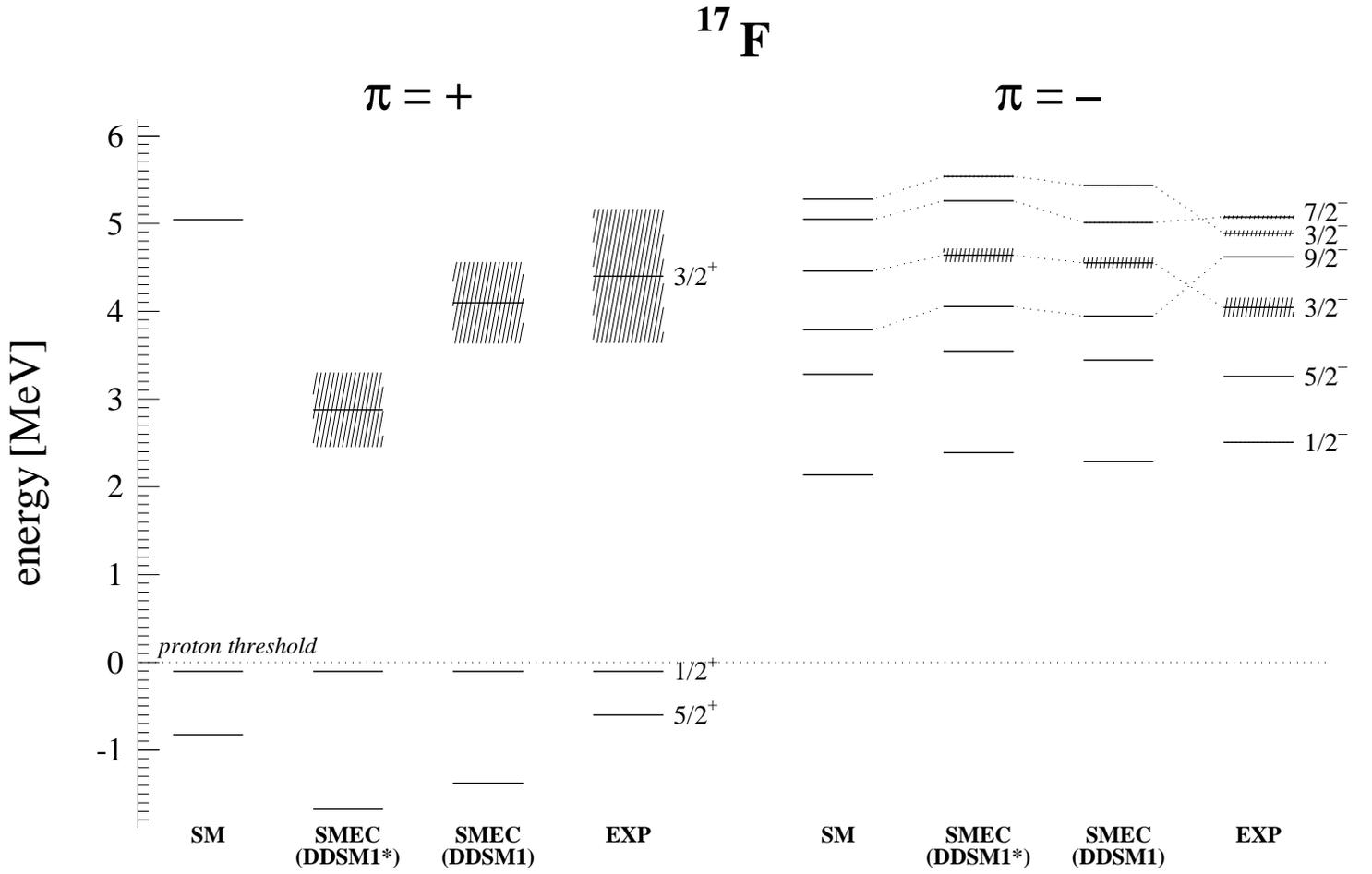,height=14cm}}
\vskip 1truecm
\caption{SM and SMEC in different approximations
labelled 'SMEC (DDSM1$^*$)' (no quenching factors for positive parity states),
SMEC (DDSM1)' (with quenching factors) vs.\ experimental states of
$^{17}\mbox{F}$ nucleus.  The proton threshold
energy is adjusted to reproduce position of the $1/2_{1}^{+}$
first excited
state. The shaded regions represent the width of resonance states. For the
details of the calculation see the description in the text and in the caption
of Table~\protect\ref{f17}.}
\label{fig17bis}
\end{figure}

Fig.~\ref{figf17} compares the SM energy spectrum of $^{17}\mbox{F}$ for
positive parity (l.h.s.\ of the plot) and negative parity (r.h.s.\ of the
plot) states, with those obtained in the SMEC approach for WB and DDSM1
residual interactions.
The experimental data are plotted separately for positive and negative
parity states as well. SM calculation
is performed using the effective interaction described in Sect.~III.A.
 In the column denoted 'SMEC (WB)' we show results of
SMEC approach with the residual coupling between $Q$ and $P$ subspaces
which is
given by the mixture of Wigner and
Bartlett forces  (\ref{force}) with the spin-exchange parameter
$(1-\alpha) =0.27$. The overall strength parameter
$V_{12}^{(0)} = 300\,$MeV$\cdot$fm$^{3}$ has been obtained
by fitting the spectra of
$^{17}\mbox{F}$ and $^{17}\mbox{O}$. In the
column denoted 'SMEC (DDSM1)', the DDSM1 density dependent residual
interaction (\ref{force1}) is used. Only the coupling matrix
elements between the $J^{\pi}=0_{1}^{+}$ g.s.\ wave function of
$^{16}\mbox{O}$ and all considered states in $^{17}\mbox{F}$ are included.
Zero of the energy scale for $^{17}\mbox{F}$ is fixed at the experimental
position of $J^{\pi}=1/2_{1}^{+}$ first excited state
with respect to the proton emission threshold
for all different examples of SMEC calculations which are
shown in Figs.~\ref{figf17} and \ref{fig17bis}.

The iterative procedure to correct $U(r)$ and  to
include the diagonal part contribution of residual interaction has been
described in the previous section.
The self-consistently determined s.p.\ potential $U^{(sc)}(r)$ is then
used to calculate radial formfactors of coupling matrix elements
and s.p.\  wave functions.  Different initial potentials $U(J^{\pi})$
(see Table~\ref{parameters})  are
used for the calculation of self-consistent
average potentials for different many-body states in $^{17}\mbox{F}$. These
different self-consistent potentials correspond to a unique equivalent SW
potential $U^{(eq)}(r)$.   The
renormalization of initial potential by the residual coupling of $Q$ and
$P$ subspaces is the same for all
states $J^{\pi}$ of the same spin and parity. For positive parity states,
as a result of the restriction in SM calculations in $Q$ subspace, the
renormalization of average potential operates only for s.p.\ orbits which
have the same spin $j$ and parity $\pi $ as those of the many body states
$J, \pi $. {\it E.g.\/}, for the  many body
state $J^{\pi} = 5/2^{+}$ of $^{17}\mbox{F}$ ($^{17}\mbox{O}$), only potential
of the $0d_{5/2}$ s.p.\ orbit is
modified by the coupling to the continuum. Similarly for
$J^{\pi} = 1/2^{+}$ many body state and the s.p.\ orbit $1s_{1/2}$.
For those s.p.\ orbits which in a given many body state
are not modified by the selfconsistent
renormalization, we calculate radial formfactors using the 'universal'
equivalent s.p.\ potential $U^{(eq)}(r)$ which does
not depend neither on the many body state nor on the s.p.\ orbit.
 Of course, this special property is only due to
a simple structure of $^{16}\mbox{O}$ and $^{17}\mbox{F}$ nuclei in our SM
calculations. For neutrons in $^{17}\mbox{F}$
there is no renormalization of the average
potential and we use the equivalent potential for neutrons to get the radial
formfactors (see Table~\ref{spen}). Supplementary informations concerning the
results shown in Fig.~\ref{figf17} can be found in Table~\ref{f17}.

The spectrum of $^{17}\mbox{F}$ is insensitive to certain
approximations in the SMEC. G.s.\ energy relative to the proton
emission threshold is reasonably well
reproduced by the SMEC. Coupling to the continuum induces strong relative
shifts of $5/2_{1}^{+}$ and $3/2_{1}^{+}$ states with respect to the
$1/2_{1}^{+}$ state and the negative parity states. The position of g.s.\ with
respect to the energy threshold for proton emission changes by about
500 keV due to the inclusion of coupling to the g.s.\ of
$^{16}\mbox{O}$ (compare columns denoted 'SM' with those denoted
'SMEC (WB)' and 'SMEC (DDSM1)'
in Fig.~\ref{figf17} and in Table~\ref{f17}). These shifts are
generally smaller for the DDSM1 residual coupling. The coupling to excited
$0_{2}^+$ state at $E^{*} = 6.049\,$MeV in $^{16}\mbox{O}$ ($0_2^{+}$
has predominantly $4p - 4h$ structure
with respect to the dominant $0p - 0h$ configuration in the g.s.\ $0_1^{+}$ )
is expected to be unimportant in $^{17}\mbox{F}$ or $^{17}\mbox{O}$.
Calculated width of states depends
on chosen residual couplings  (compare 'SMEC (WB)' with 'SMEC (DDSM1)') and in
general the agreement is better for the density dependent interaction.
The effective quenching of the coupling matrix elements
(\ref{ef4a}--\ref{ef4c}) for positive parity states does not solve this
problem completely. To get a better agreement with the data,
the energy separation
of $5/2_{1}^{+}$ and $1/2_{1}^{+}$ SM states should be decreased while
the splitting of $5/2_{1}^{+}$ and $3/2_{1}^{+}$ SM states should remain
unchanged.
This means that the separation of $1s$ subshell and the centroid od $0d$
subshells should be increased by few hundred keV. We did not however pursue the
studies  in this direction  and we keep standard values of s.p.\ energies
\cite{gemo}.

The effect of quenching of the coupling operator between $Q$ and $P$ can be
seen in Fig.~\ref{fig17bis} where the comparison between the SMEC
calculations with the quenching factors (entry 'SMEC (DDSM1)') and those
without quenching factors (entry 'SMEC (DDSM1$^{*}$') are
shown. The strong downshift due to the continuum
coupling is seen for the ($5/2_1^+$ - $3/2_1^+$) - pair with respect
to both the negative parity states and
the $1/2_1^{+}$ state which moves downwards much less.
The effect of quenching of the coupling operator on negative parity states is,
as expected, very weak.

A useful measure of the radial wave function is the $B(E2)$ transition matrix
element between $1/2_1^+$ and $5/2_1^+$ bound states. In SMEC,
we obtain 70.3 e$^2$fm$^4$ for DDSM1 residual coupling. This value has
been calculated using the effective charges :
$e_p = 1.41$ and $e_n = 0.47$ and the BG quenching factors
(Table~\ref{decompo2}). The experimental value for this transition is
$B(E2)_{exp} = 64.92\,$e$^2$fm$^4$. The SM prediction for this transition is
$B(E2)_{SM} = 32.7\,$e$^2$fm$^4$ assuming the oscillator length
$b = A^{1/6}\,$fm and the same
effective charges as used in the SMEC calculations. The better agreement of
calculated in SMEC
$B(E2)$ values with the data is mainly due to more realistic radial
dependence of the $1s_{1/2}$ s.p.\ orbit in $J^{\pi} = 1/2_1^{+}$ many body
state. The rms radius for this orbit in SMEC
is ${<r^2>}^{1/2} = 5.212\,$fm, as compared to the value of
${<r^2>}^{1/2} = 3.629\,$fm for the $0d_{5/2}$ s.p.\ orbit in
$J_1^{\pi} = 5/2^{+}$ state. Nuclear rms radius of
$^{17}$F is 2.664~fm in the g.s.\ and 2.814~fm in the first excited state
$1/2^+$.

\subsubsection{Spectrum of $^{17}\mbox{O}$}
\begin{figure}[t]
\centerline{\epsfig{figure=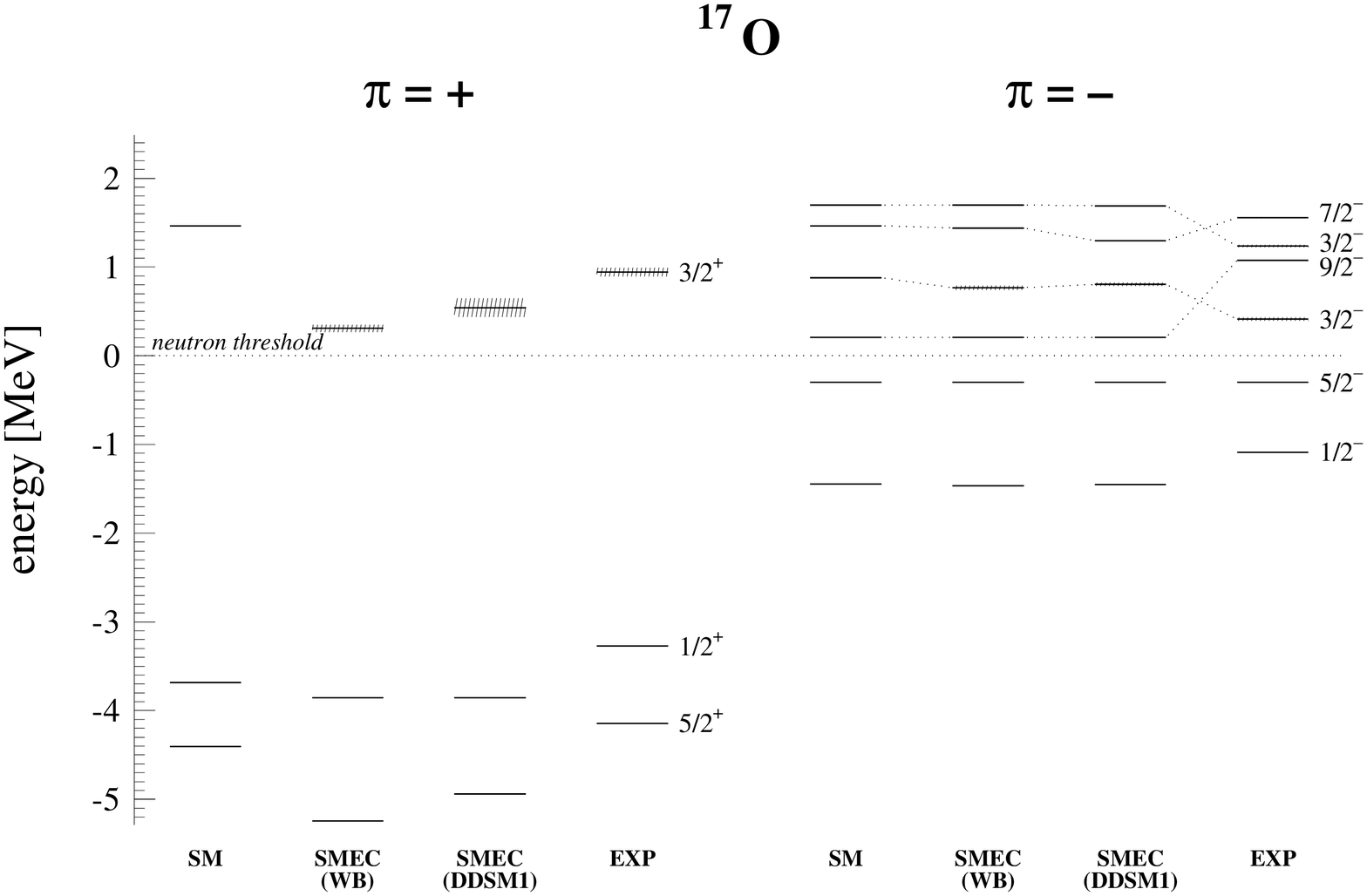,height=14cm}}
\vskip 1truecm
\caption{The same as in Fig.~\protect\ref{figf17} but for
$^{17}\mbox{O}$ nucleus.  The proton threshold
energy is adjusted to reproduce position of the $5/2_{1}^{-}$
excited
state. The shaded regions represent the width of resonance states. For the
details of the calculation see the description in the text and in the caption
of Table~\protect\ref{o17}.}
\label{figo17}
\end{figure}

Fig.~\ref{figo17} compares the SM energy spectrum of $^{17}\mbox{O}$ for
positive parity (l.h.s.\ of the plot) and negative parity (r.h.s.\ of the
plot) states, with those obtained in the SMEC approach for WB and DDSM1
residual interactions.
The experimental data are plotted separately for positive and negative
parity states as well. SM calculation
is performed using the same effective interaction as used for $^{17}\mbox{F}$
spectrum shown in Figs.~\ref{figf17} and \ref{fig17bis}.
 In the column denoted 'SMEC (WB)' we show results of
SMEC approach with the residual coupling given by the mixture of Wigner
and Bartlett forces  (\ref{force}). In the column
denoted 'SMEC (DDSM1)', the density dependent interaction DDSM1 is used as
the residual interaction. Parameters of the residual couplings are
the same as used to describe the $^{17}\mbox{F}$ spectrum.
Only the coupling matrix
elements between the $J^{\pi}=0_{1}^{+}$ g.s.\ wave function of
$^{16}\mbox{O}$ and all considered states in $^{17}\mbox{O}$ are included.
For all different examples of SMEC calculations which are
shown in Fig.~\ref{figo17},
zero on the energy scale is fixed at the experimental
position of $J^{\pi}=5/2_{1}^{-}$ state which is close to the neutron
emission threshold.

The self-consistently determined s.p.\ potential $U^{(sc)}(r)$ is then
used to calculate radial formfactors of coupling matrix elements
and s.p.\  wave functions.  Different initial potentials $U(J^{\pi})$ for
[n $\bigotimes$ $^{16}$O] (see Table~\ref{parameters})  are
used for the calculation of self-consistent
average potentials for different many-body states in $^{17}\mbox{O}$. These
different self-consistent potentials correspond to a unique equivalent SW
potential $U^{(eq)}(r)$ (see Table~\ref{spen1}).  For protons
there is no correction from the residual interaction modifying the average
potential and we use the equivalent potential for protons to get the radial
formfactors (see Table~\ref{spen1}).

The energy intervals between the g.s.\ and the
positive parity states $1/2_1^+$ and $3/2_1^+$ in SMEC(DDSM1) calculations
perfectly reproduce the experimental energy sequence for these levels. The WB
residual force leads to a too strong relative shift of $5/2_1^+$ with respect
to $1/2_1^+$  and $3/2_1^+$  states.
The quenching factors have been introduced in the
calculations for positive parity states similarly as discussed before for
$^{17}\mbox{F}$. One should remind that in $^{17}\mbox{F}$, in spite of using
the quenching factors, the
$5/2_1^+$ state was shifted downwards with respect to other
positive parity states. Improved situation in $^{17}\mbox{O}$ is due to the
larger separation of this state with respect to the particle continuum.
Nevertheless, the coupling to the continuum is necessary here to bring closer
the $3/2_1^+$  and  $1/2_1^+$ SM states.
The shift of the ground state
energy in SMEC relative to the neutron emission threshold is solely due to the
choice of zero on the energy scale at the experimental position of $5/2_1^-$
weakly bound state. Calculated width of states depends
on chosen residual couplings  (compare 'SMEC (WB)' with 'SMEC (DDSM1)') and in
general the agreement is better for the density dependent interaction.

As a measure of the radial wave function in $^{17}\mbox{O}$ we calculate
the $B(E2)$ transition matrix
element between $1/2_1^+$ and $5/2_1^+$ bound states. In the SMEC,
we obtain $5.1\,$e$^2$fm$^4$ for DDSM1 residual coupling. Changing
the residual
coupling and using WB interactions leaves this value almost unchanged.
The $B(E2)$ value in SMEC has
been calculated using the effective charges:
$e_p = 1.41$ and $e_n = 0.47$, and the BG quenching factors
(Table~\ref{decompo2}). The experimental value for this transition is
$B(E2)_{exp} = 6.2\,$e$^2$fm$^4$ whereas the SM prediction is
$B(E2)_{SM} = 3.62\,$e$^2$fm$^4$ assuming the oscillator length
$b = A^{1/6}\,$fm and taking the same
effective charges as in the SMEC calculations. The close agreement of
calculated in SMEC $B(E2)$ values with the data is due to a
more realistic radial
dependence of the $1s_{1/2}$ s.p.\ orbit in $J^{\pi} = 1/2_1^{+}$ many body
state.  The rms radius for this orbit
is ${<r^2>}^{1/2} = 4.184\,$fm, as compared to the value:
${<r^2>}^{1/2} = 3.439\,$fm for the $0d_{5/2}$ s.p.\ orbit in
$J_1^{\pi} = 5/2^{+}$ many body  state.

\section{ 
THE  ASTROPHYSICAL FACTOR FOR
$^{16}\mbox{O}(p,\gamma )^{17}\mbox{F}$ }

\begin{figure}[t]
\centerline{\epsfig{figure=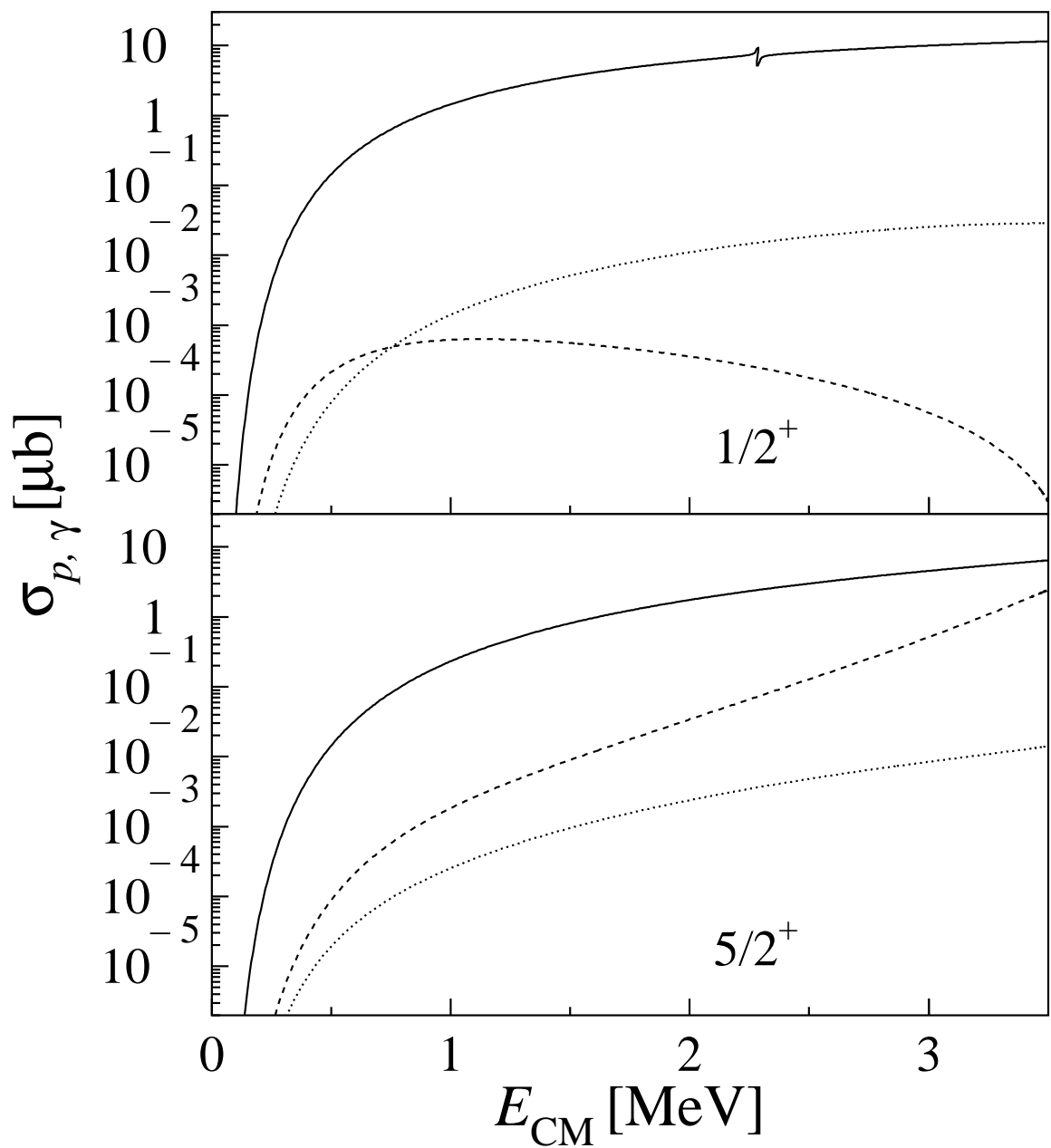,height=18cm}}
\vskip 1truecm
\caption{Multipole contributions to the capture cross section to the
g.s.\ $^{16}\mbox{O}(p,\gamma)^{17}\mbox{F}(J^{\pi}={\frac{5}{2}}^{+})$ and
to the first excited state
$^{16}\mbox{O}(p,\gamma)^{17}\mbox{F}(J^{\pi}={\frac{1}{2}}^{+})$ are plotted
as a function of the center of mass energy $E_{CM}$. The SMEC calculations
for $E1$ (the solid line), $E2$ (the dotted line) and $M1$ (the dashed line)
have been done for
the DDSM1 residual interaction (\protect\ref{force1}).
For more details see the description in the text.}
\label{figadd}
\end{figure}

In Fig.~\ref{figadd} we show the calculated multipole contributions
($E1$, $E2$, $M1$) to the total capture cross section as a function of the
c.m.\ energy $E_{CM}$, separately for the transitions to the
g.s.\ $5/2_{1}^{+}$  and to the first excited
state $1/2_{1}^{+}$ in $^{17}\mbox{F}$.
The SMEC calculation is done with the DDSM1 residual
interaction as used in the calculations of spectra shown in Figs.~3 and 4.
Parameters of initial potentials $U(5/2^{+})$ and $U(1/2^{+})$
can be read from Table~\ref{parameters}.
The proton threshold energy is adjusted to agree energies of calculated and
experimental $1/2_{1}^{+}$ state with respect to the proton emission threshold.
 The photon energy is then given by the difference of
c.m.\ energy of $[^{16}\mbox{O} + \mbox{p}]_{J_{i}}$~ system
and the experimental energy of the final state $J_f$ in $^{17}\mbox{F}$.

The dominant contribution to the total capture cross-section
for both $5/2_1^{+}$
and $1/2_1^{+}$ final states, is provided by $E1$ transitions from the
incoming $p$ wave to the bound $0d_{5/2}$ and $1s_{1/2}$ states. We took into
account all possible $E1$, $E2$, and $M1$ transitions from incoming $s$,
$p$, $d$, $f$, and $g$ waves but only $E1$ from incoming $p$ - waves
give important contributions. In the transition to the g.s., the $E1$
contribution from incoming $f_{7/2}$ wave gives is by
a factor $\sim$100 smaller than the contribution from $p_{3/2}$ wave
for energies of incoming proton up to 3.5~MeV.
The $E2$ contribution is smaller by at least
three orders of magnitude in both branches.

In the transition to
the first excited state $1/2_1^+$, the dominant
$M1$ contribution comes from the incoming $s_{1/2}$ wave and is by $\sim$2
orders of magnitude smaller than the $E1$ contribution.
The $M1$ contribution is important, in particular,
in the branch $^{16}\mbox{O}(p,\gamma)^{17}\mbox{F}(J^{\pi}=5/2_1^{+})$ for
$E_{CM} > 3\,$MeV, where its contribution is of a similar order as the $E1$
contribution due to $3/2_1^+$ resonance in the $d_{3/2}$ scattering wave.
One should also notice that the energy dependence of $M1$
component in the capture cross section is totally different
in $^{16}\mbox{O}(p,\gamma)^{17}\mbox{F}(J^{\pi}=1/2_1^{+})$ and
$^{16}\mbox{O}(p,\gamma)^{17}\mbox{F}(J^{\pi}=5/2_1^{+})$ branches.
The contribution of $M1$ for
$^{16}\mbox{O}(p,\gamma)^{17}\mbox{F}(J^{\pi}=1/2_1^{+})$ is decreasing
with increasing energy above $E_{CM} \sim 1\,$MeV and approaches zero for
$E_{CM} \sim 3.5\,$MeV. The $M1$ contribution begins to grow again at
higher energies.
Since ${\sigma}_{M1}$ cross section for $1/2_1^+$ final state originates only
from the absorption of $s$-wave proton and, moreover, since by construction:
${<\Psi_f|\Psi_i>} = 1$ (the spectroscopic factor of $1/2_1^+$ state
equals 1), therefore the energy behavior
of ${\sigma}_{M1}$  cross section is governed by the
Breit--Wigner like factor coming from the $1/2_1^+$
many-body wave function (see eq. (\ref{eq2}) or (\ref{cons})). So
${\sigma}_{M1}$ depends
crucially on the energy dependence of this single eigenvalue.
While the real part of $E(J^{\pi} = 1/2_1^+)$ eigenvalue
behaves smoothly with energy, the
imaginary part, on the contrary,  after initial rise
becomes almost equal to zero at $E_{CM} \simeq 3.5\,$MeV
and starts to rise again for higher energies. Since we have
only one $J^{\pi} = 1/2^+$ state in our SM space,
the energy behavior of its eigenvalue determines directly the behavior of
${\sigma}_{M1}$ cross-section,
because in the limit: $\Gamma \rightarrow 0$, the Breit--Wigner factor
behaves like the $\delta$-function. In the more complete SM calculations
which yield more $J^{\pi} = 1/2^+$ states, this unusual effect is expected
to be reduced. Nevertheless, because its principal cause is the
single-partial-wave characteristic of this transition, we believe that the
trace of it remains.

In the branch
$^{16}\mbox{O}(p,\gamma)^{17}\mbox{F}(J^{\pi}=1/2_1^{+})$, one may also
notice small resonant part in $E1$ at the position of $1/2_1^{-}$ resonance,
but otherwise the cross-section has non-resonant
behavior. The non-resonant $E1$ contribution is a good measure of spatial
extension of $5/2_{1}^{+}$ and $1/2_{1}^{+}$ wave functions respectively. For
the decay branch $^{16}\mbox{O}(p,\gamma)^{17}\mbox{F}(J^{\pi}=1/2_1^{+})$,
due to the very small binding energy of $J^{\pi}=1/2_{1}^{+}$ state ($Q =
105$ keV) and its particularly
simple structure, this cross section is a sensible measure of the
extension of the proton $1s_{1/2}$ orbit. It is essential for the calculated
cross-section that the
$1s_{1/2}$ proton orbit in the self-consistent average potential for
$J^{\pi}=1/2^+$ is bound by
$105\,$keV. Even small modification of
this value by different choice of the depth parameter $V_0$ in $U(r)$,
introduce the modification of $S_{\mbox{\scriptsize E1}}$ which is larger
than any modification due to possible uncertainties
in the potential radius $R_0$ or its surface diffuseness $a$.

\begin{figure}[t]
\centerline{\epsfig{figure=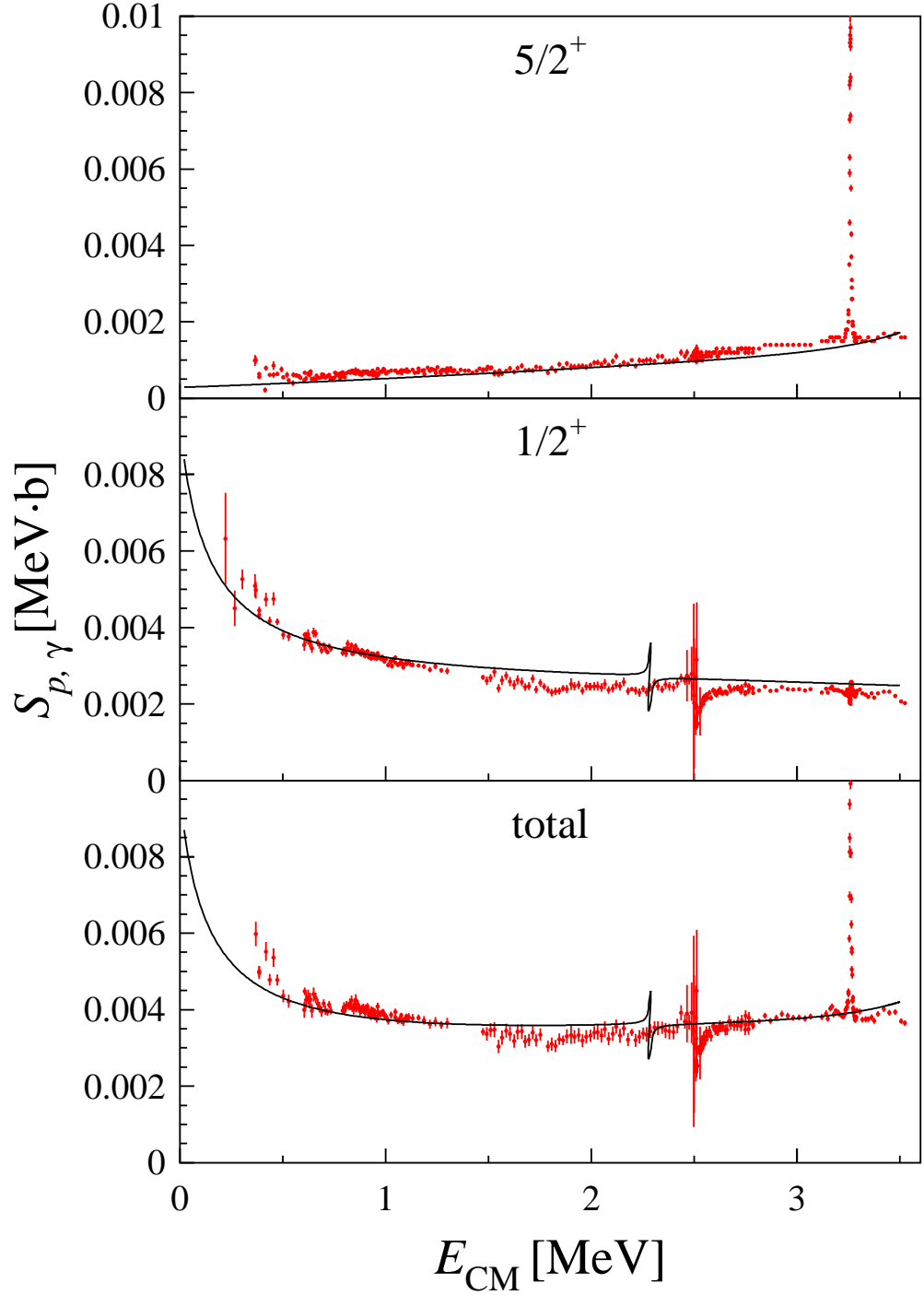,height=20cm}}
\vskip 1truecm
\caption{The astrophysical $S$-factor for the reactions
$^{16}\mbox{O}(p,\gamma)^{17}\mbox{F}(J^{\pi}=5/2_{1}^{+})$ and
$^{16}\mbox{O}(p,\gamma)^{17}\mbox{F}(J^{\pi}=1/2_{1}^{+})$ is plotted
as a function of the center of mass energy $E_{CM}$. The experimental data are
from \protect\cite{morlock}. The SMEC calculations
(the solid line) have been done for
the DDSM1 residual interaction (\protect\ref{force1}).
For more details see the description in the text.}
\label{figa}
\end{figure}

Fig.~\ref{figa} shows the total $S$-factor
as a function of the c.m.\ energy, as well as its values for the
$^{16}\mbox{O}(p,\gamma)^{17}\mbox{F}(J^{\pi}=1/2_1^{+})$ and
$^{16}\mbox{O}(p,\gamma)^{17}\mbox{F}(J^{\pi}=5/2_1^{+})$ branches. The overall
agreement with the experimental data of Morlock et al \cite{morlock}
is very good. The resonance contributions of $1/2_{1}^{-}$ in
$^{16}\mbox{O}(p,\gamma)^{17}\mbox{F}(J^{\pi}=1/2_1^{+})$
is well reproduced even
though their precise energy position is slightly shifted. This resonance
contributes only through its $P$ subspace continuation because the
spectroscopic
factor for this state is equal zero. The contribution of $5/2_1^{-}$
resonance in  the
$^{16}\mbox{O}(p,\gamma)^{17}\mbox{F}(J^{\pi}=5/2_1^{+})$ branch is so
narrow in
energy that has been omitted in the figure. Energy of this resonance
in SMEC (DDSM1) is $E_{CM}=3.44\,$MeV.

\begin{figure}[t]
\centerline{\epsfig{figure=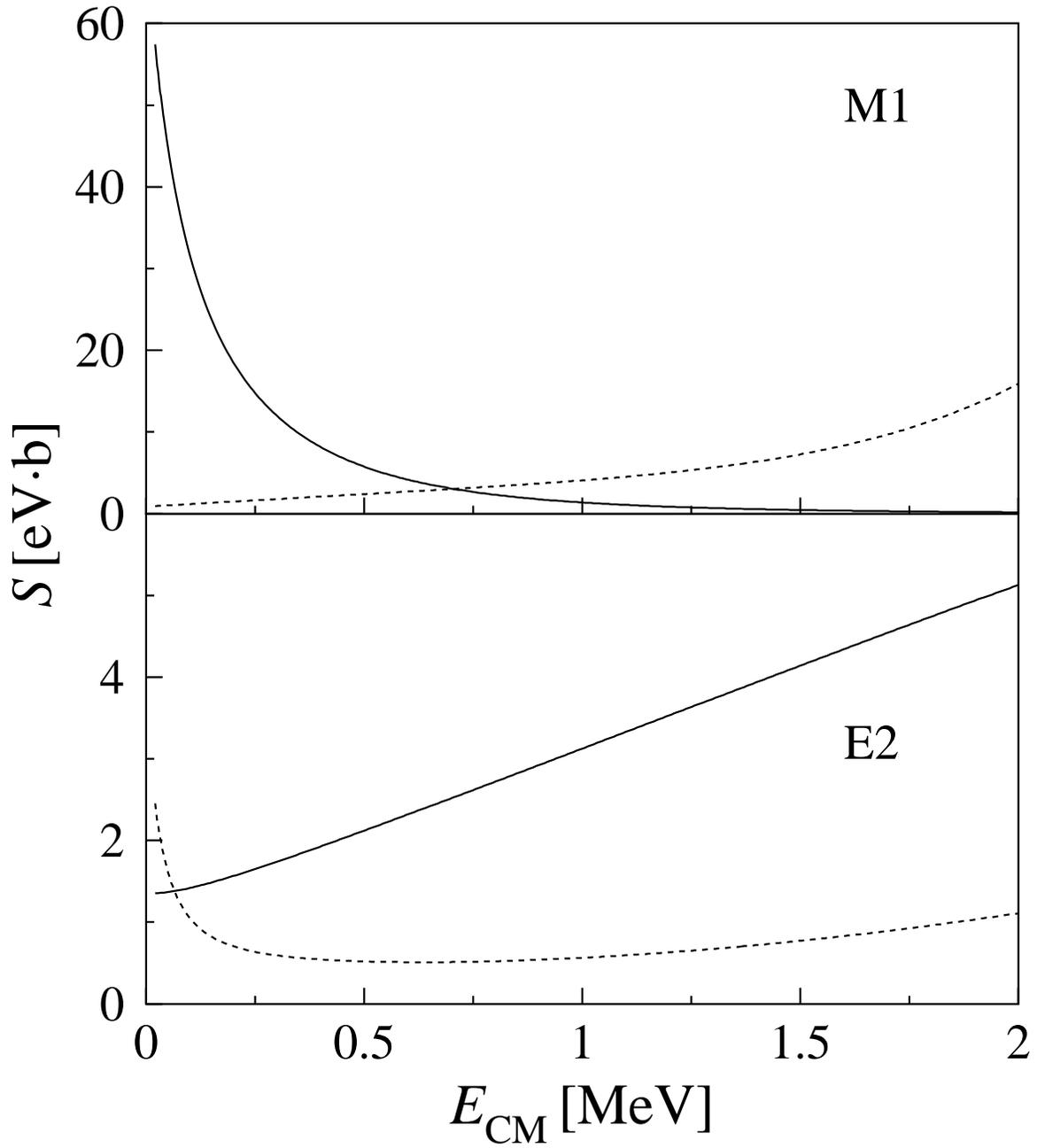,height=18cm}}
\vskip 1truecm
\caption{The $E2$ and $M1$ components of the
astrophysical $S$-factor for the reactions
$^{16}\mbox{O}(p,\gamma)^{17}\mbox{F}(J^{\pi}=5/2_{1}^{+})$ (the dashed lines)
and  $^{16}\mbox{O}(p,\gamma)^{17}\mbox{F}(J^{\pi}=1/2_{1}^{+})$ (the solid
lines) is plotted as a function of the center of mass energy $E_{CM}$.
 The SMEC calculations have been done for
the DDSM1 residual interaction (\protect\ref{force1}).
For more details see the description in the text.}
\label{fign}
\end{figure}

The strong increase of the $S$-factor at low energies for $1/2^{+}$
component is a direct consequence of the proximity of continuum for
$1s_{1/2}$ s.p.\ orbit in $J^{\pi}=1/2_{1}^{+}$ state. The energy dependence
of $S$-factor as $E_{CM} \rightarrow 0$ has been fitted by a  second
order polynomial to theoretical points obtained in the interval from 13 to
50~keV in steps of 1~keV. For the DDSM1 residual coupling we
have $S(0) = 9.25\times 10^{-3}$~MeV$\cdot$b, and the logarithmic
derivative is
$S^{'}(0)/S(0) = -4.98$~MeV$^{-1}$.  The ratio of M1 to E1
transition is $S_{\mbox{\scriptsize M1}}(0)/S_{\mbox{\scriptsize E1}}(0) =
7.55\times 10^{-3}$. The $E1$, $M1$, $E2$ contributions from different incoming
waves at $E_{CM} \rightarrow 0$ can be seen in Table~\ref{du1}. The $E2$ and
$M1$ components of the $S$-factor for the branches
$^{16}\mbox{O}(p,\gamma)^{17}\mbox{F}(J^{\pi}=1/2_1^{+})$ and
$^{16}\mbox{O}(p,\gamma)^{17}\mbox{F}(J^{\pi}=5/2_1^{+})$ summed over all
partial incoming wave contributions are shown in Fig.~\ref{fign}. For the WB
residual coupling, $E1$ and $E2$ contributions are only $\sim 0.7$\%
smaller but M1 is about 20 times
bigger than for the DDSM1 case, so the ratio of M1 to E1
transition is $S_{\mbox{\scriptsize M1}}(0)/S_{\mbox{\scriptsize E1}}(0) =
0.154$.  The ratio of $E2$ and $E1$ contributions for
$^{16}\mbox{O}(p,\gamma)^{17}\mbox{F}(J^{\pi}=1/2_1^{+})$ is:
${\sigma}^{E2}/{\sigma}^{E1}=1.62\times 10^{-4}$, $2.22\times 10^{-4}$ and
$5.425\times 10^{-4}$~ at 20, 100 and 500~keV, respectively.

For the deexcitation to the g.s.\ $5/2_1^+$, the fit of calculated
$S$-factor as  $E_{CM} \rightarrow 0$ yields:
$S(0)=2.89 \times 10^{-4}$~MeV$\cdot$b and
$S'(0)/S(0) = 0.767$~MeV$^{-1}$.  The ratio of $M1$ to $E1$ transition is
$S_{\mbox{\scriptsize M1}}(0)/S_{\mbox{\scriptsize E1}}(0) =
3.1\times 10^{-3}$.
$E1$, $M1$, $E2$ contributions from different incoming
waves at $E_{CM} \rightarrow 0$ can be seen in Table~\ref{du2}. It is
particularly interesting to notice large $E2$ contribution from
$s_{1/2}$ incoming wave having large negative logarithmic derivative.
The ratio of $E2$ and $E1$ contributions for
$^{16}\mbox{O}(p,\gamma)^{17}\mbox{F}(J^{\pi}=5/2_1^{+})$ is:
${\sigma}^{E2}/{\sigma}^{E1}=8.5\times 10^{-3}$, $3.5\times 10^{-3}$ and
$1.35\times 10^{-3}$~ at 20, 100 and 500~keV, respectively.  In the total
cross
section this ratio is: ${\sigma}^{E2}/{\sigma}^{E1}=4.4\times 10^{-4}$,
$3.7\times 10^{-4}$ and $6.1\times 10^{-4}$~ at 20, 100 and 500~keV,
respectively.

\section{Elastic cross-section and phase-shifts}
\begin{figure}[t]
\centerline{\epsfig{figure=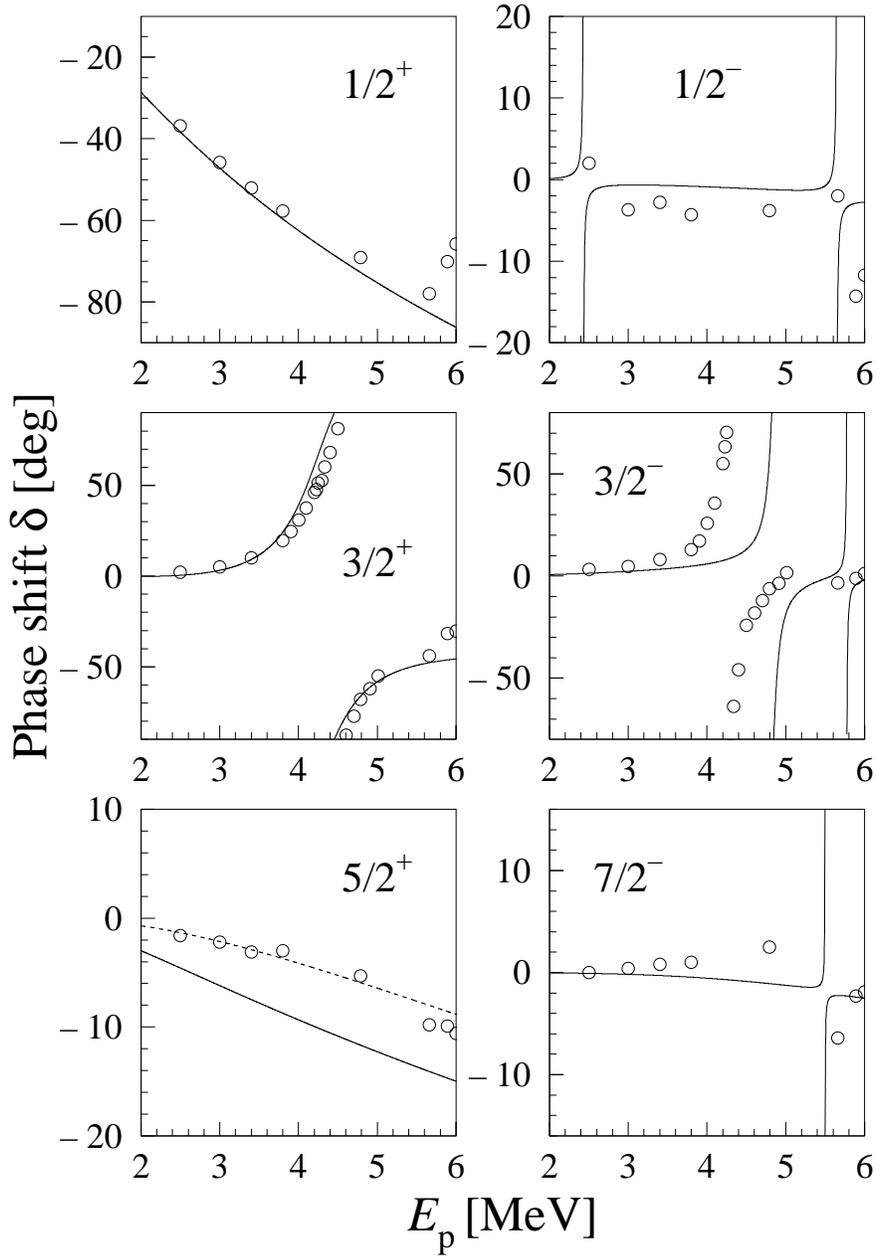,height=18cm}}
\vskip 1truecm
\caption{The phase-shifts for the $p + ^{16}\mbox{O}$ elastic scattering
as a function of the proton energy
$E_{p}$ for the ${\frac{1}{2}}^{+}$,
${\frac{3}{2}}^{+}$,  ${\frac{5}{2}}^{+}$, and  ${\frac{1}{2}}^{-}$,
${\frac{3}{2}}^{-}$,  ${\frac{7}{2}}^{-}$  partial waves.
The experimental data are from \protect\cite{bluehab}.
SMEC results have been obtained for the DDSM1 residual interaction
(\protect\ref{force1}) and are plotted with the solid line. For
${\frac{5}{2}}^{+}$ partial wave, we show also the SMEC results without the
resonant part (the dashed line). For more information see the description in
the text.}
\label{figc_1}
\end{figure}

\begin{figure}[t]
\centerline{\epsfig{figure=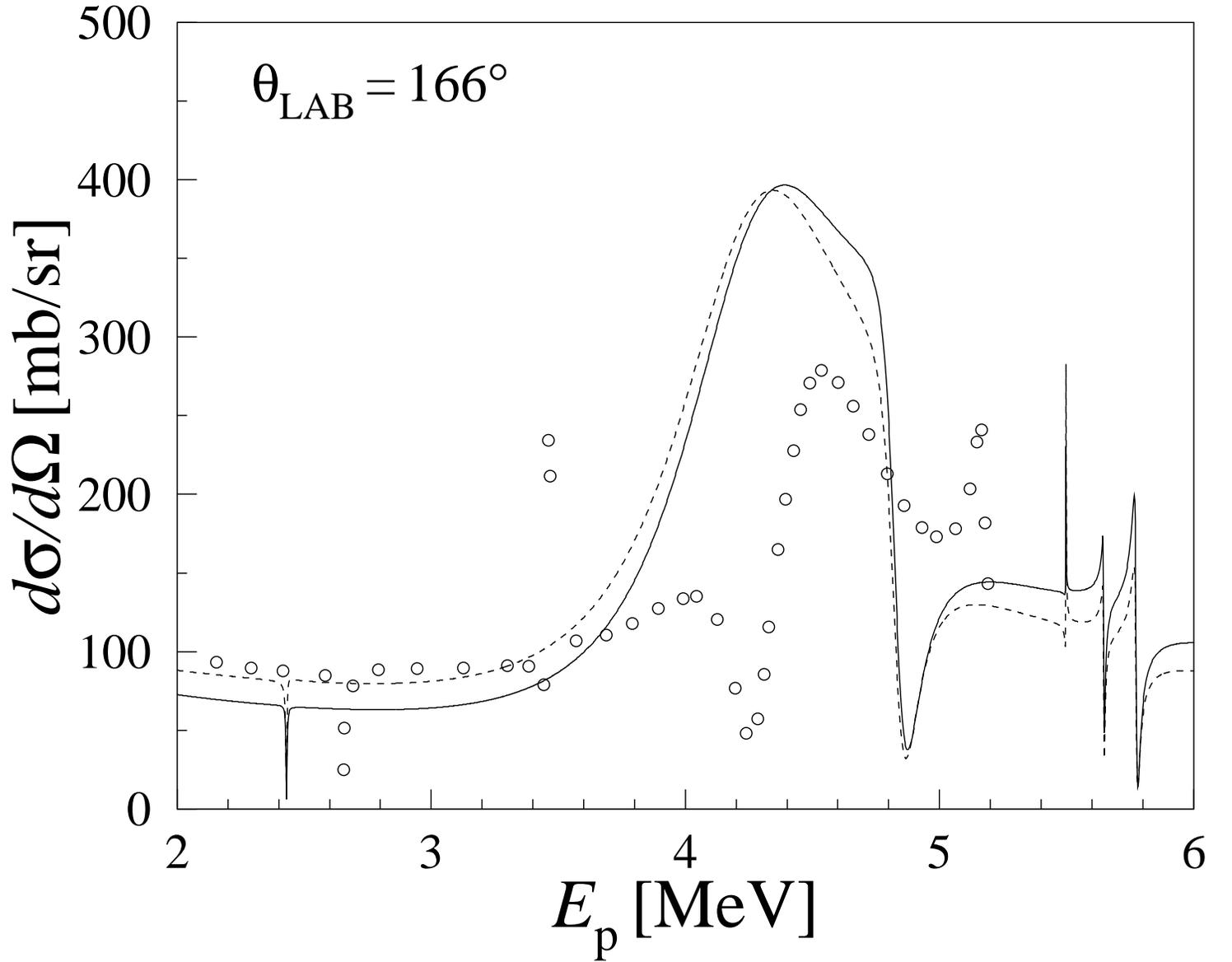,height=18cm}}
\vskip 1truecm
\caption{The elastic cross-section at a laboratory angle $\theta_{LAB} = 166
^\circ$ for the $p + {^{16}\mbox{O}}$ scattering as a function of proton
energy $E_p$. The SMEC calculations (the solid line)
have been performed using the DDSM1
residual interaction. The results shown with the dashed line do not include
the resonant contribution of $5/2_1^+$ state. Experimental
cross-sections are from Ref.~\protect\cite{salisbury}.}
\label{figb}
\end{figure}

Another observable quantities that can be calculated using the solutions of
the
SMEC are the elastic phase shifts and elastic cross-sections for different
proton bombarding energies. Results for the elastic phase shifts,
shown in Figs.~\ref{figc_1}, and those for the elastic cross-section, shown
in Fig.~\ref{figb}, have been obtained using the SMEC solutions for identical
parameters of initial potentials and the DDSM1 residual coupling between $Q$
and $P$ subspaces as discussed before
in Sect.~III.D.1 for the spectra of $^{17}\mbox{F}$
and in Sect.~IV for the capture reaction
$^{16}\mbox{O}(p,\gamma)^{17}\mbox{F}(J^{\pi})$. The elastic phase shifts are
well reproduced by the SMEC for all states except for the $5/2^+$. In this case
the calculated and experimental \cite{bluehab} values differ by
about 5$^\circ$.
This discrepancy disappears when the resonant contribution to this phase
shift is removed from the SMEC solution (compare the solid and dashed lines in
the plot for $5/2^+$ in Fig.~\ref{figc_1}). It may be
surprising at first sight to see so large resonant contribution to the
phase shift associated with the g.s.\ wave function.
For energies below the proton emission threshold the coupling to
the continuum introduces only the hermitean modifications of the Hamiltonian
which shift the energy of $5/2_1^+$ state in SMEC with respect to its initial
position given by the SM but do not generate any width for this state.
The coupling of $Q$ and $P$ subspaces is non-local and, hence, the
effective Hamiltonian in SMEC is energy dependent. For excitation energies
above the proton threshold, the coupling of $Q$ and $P$ subspaces
generates the non-hermitean
corrections to the effective Hamiltonian which yield the imaginary part
of the eigenvalue and generates the resonant like behavior which is so well
visible in the $5/2_1^+$ elastic phase shift. The origin of this large 'halo'
of $5/2_1^+$ bound state
in the continuum for positive energies and the large shift
of the real part of the $5/2_1^+$ eigenvalue for negative energies (see Fig.
\ref{figf17} and \ref{fig17bis}) is the same and should be traced back
to the incorrect description of correlations in the SM wave function for
this state. As discussed in the previous chapter, one
expects that the $5/2_1^+$ state in $^{17}\mbox{F}$ and
$^{17}\mbox{O}$ is not a pure s.p.\ excitation outside of the $^{16}\mbox{O}$
core \cite{brown}. Therefore, it is also natural to quench the matrix
elements of $Q$ - $P$ coupling as described in Sect.~III.B. This
procedure, which consists of using quenching factors for positive
parity states to correct for missing higher order $\mbox{n}p - \mbox{n}h$
components in their wave function, is actually not sufficient as
the example of elastic phase shift for $5/2_1^+$ demonstrates clearly.
The quenching procedure cures the problem of unphysically large energy shift
for the $5/2_1^+$ state due to the coupling to the continuum
but does not correct the
wave function for missing correlations which in turn lead to the disappearance
of 'halo' of this state for positive energies. This deficiency of present SM
calculations in $^{17}\mbox{F}$ and $^{17}\mbox{O}$, which is seen
consistently both in the energy spectrum as well as in the elastic phase shifts,
demonstrates  potentiality of SMEC approach for mass-regions far off
the $\beta$ - stability valley where the experimental information
about exotic nuclei will be scarce and one will
have to use both the spectroscopic information and the  information from the
scattering experiments to learn about the structure of those nuclei.
This example shows also that in the
SMEC approach which unifies description of discrete state properties and the
scattering continuum, one
may use different kinds of experimental
data to fix those few parameters of the model such as the
overall strength of the residual $Q$ - $P$ coupling or the radius and
diffuseness of the initial average potential.

Elastic excitation functions at a laboratory angle of 166$^\circ $ calculated
in
SMEC with DDSM1 residual interaction are compared with the experimental data
\cite{salisbury} in Fig.~\ref{figb}. The calculated values at low energies
are too low as compared to
the data and this is again due to the too strong halo of $5/2_1^+$
state for positive energies (compare the dashed and solid lines in Fig.
\ref{figb} at low energies). Removal of the resonant contribution from this
state to the elastic cross section brings the calculations close to the data.
The agreement between calculated and experimental low energy cross-sections
provides a supplementary check of the spatial extension of
self-consistent potential and, hence, of the radial form factors of
s.p.\ wave functions. One should however keep in mind that, in general, the
information from the elastic
cross-sections may be strongly perturbed by the non-locality effects in the
SMEC effective Hamiltonian which
depend strongly on the many-body correlations in the SM wave functions as the
above example demonstrates.

On the average, the agreement between experimental and SMEC results for
the elastic excitation functions is reasonable if one keeps in mind large
sensitivity of this experimental measure to even small inaccuracies in the
energy position and width of resonances. The model predicts
correctly the interference pattern due to $1/2_1^-$ resonance
at $E_p \sim 2.6$ MeV and $3/2_2^-$ resonance at $E_p \sim 5.2$ MeV. Also, the
$5/2_2^-$ resonance is correctly predicted by the calculations though its
width is too narrow to be presented in the figure. For the same reason, this
resonance was not plotted in the proton capture cross-section for the
$^{16}\mbox{O}(p,\gamma)^{17}\mbox{F}(J^{\pi}=5/2_1^{+})$ branch in
Fig.~\ref{figa}. The
difference between interference patterns in the data and in the SMEC calculation
for 3.8 MeV $< E_p <$ 5 MeV is mainly due to the reversed order of
$3/2_1^-$ and $3/2_1^+$ resonances in
SMEC calculations  as compared to the data. Consequently, the
hole in the experimental elastic excitation function at $E_p \sim 4.2$ MeV,
due to $3/2_1^-$ resonance interfering with the  $3/2_1^+$ resonance cuts
sharply only the high energy tail of the $3/2_1^+$ resonance.

\section{Summary and Outlook}
In this work we have applied the SMEC approach
for the microscopic description of $^{17}\mbox{F}$ and $^{17}\mbox{O}$
spectra, the
low-energy radiative capture cross sections in the reaction
$^{16}\mbox{O}(p,\gamma)^{17}\mbox{F}$, and the elastic cross section for the
reaction
$^{16}\mbox{O}(p,p)^{16}\mbox{O}$. In the SMEC model, which
is a development of CSM model \cite{bartz1,bartz2} for the description of low
energy properties of weakly bound nuclei,
realistic SM solutions for (quasi-)bound states are coupled to the
one-particle scattering continuum. For that reason, we use
realistic SM effective interaction in the $Q$ subspace and introduce
residual force which couples $Q$ and $P$ subspaces. (For this residual
coupling we take either a combination of Wigner and Bartlett forces or
the density dependent DDSM1 interaction which is similar to the
Landau - Migdal type of interactions.)
This deliberate choice of interactions implies that
the finite-depth potential generating $P$ space and matrix elements of
the residual $Q - P$ coupling, have to be determined self-consistently.
The self-consistent iterative procedure yields then
new state-dependent average potentials and consistent with them
new renormalized matrix elements of the
coupling force. These renormalized couplings and average potentials
are then consistently used both in $Q$ and $P$
subspaces for the calculations of spectra, capture cross-sections,
elastic cross-sections, elastic phase-shifts, etc.

Simultaneous studies of spectroscopy in mirror systems as well as different
reactions involving one nucleon in the continuum, allow for a better
understanding of the role of different approximations and parameters
in the model. The dependence on radius, diffuseness or spin-orbit coupling
parameters of the initial potential $U(r)$ is not very
important and they can be taken from any reasonable systematics.
On the contrary, the depth of $U(r)$ has
to be carefully adjusted so that the energies of s.p.\ orbits in $U^{(sc)}(r)$
for $[{n} \bigotimes (N-1)]$ and $[{p} \bigotimes (N-1)]$ systems, whenever
their identification is possible,
reflect the binding of many-body states near the particle emission threshold
in the nucleus $N$. This is very important for the quantitative description of
reaction cross-sections. In the case of
$^{17}\mbox{F}$ ($^{17}\mbox{O}$), correct identification of $1s_{1/2}$
and $0d_{5/2}$ s.p.\ orbits and hence the determination
of an appropriate depth parameter in
$U(1/2^{+})$ and $U(5/2^{+})$ is unambiguous  because the spectroscopic
amplitudes in $1/2_{1}^{+}$ and  $5/2_{1}^{+}$ are close to 1.
Different binding of mirror nuclei $^{17}\mbox{F}$/$^{17}\mbox{O}$ leads
for the same $J^{\pi}$ many-body states to different $U^{(sc)}(r)$
in these nuclei. This in turn causes breaking of mirror symmetry of SM
spectra for these nuclei and is, {\it e.g.},  an essential ingredient in
understanding the difference between $B(E2)$ values for the
transition $5/2_1^{+} \rightarrow 1/2_1^{+}$ in $^{17}\mbox{F}$
 and $^{17}\mbox{O}$, as discussed in Sect.~III.

SMEC model in its present form includes the
coupling to one-nucleon continuum. The wealth of experimental data can be
described in a unified framework of SMEC in this approximation. These
include: (i) the calculation of energy spectra, $B(\Pi \lambda)$ transition
matrix elements and various static nuclear moments such as the magnetic
or mass/charge quadrupole moments etc., (ii) the calculation of various
radiative capture processes: $(p, \gamma )$, $(n, \gamma )$, Coulomb
breakup processes: $(\gamma , p)$,  $(\gamma , n)$  and
 elastic or inelastic cross sections $(p,p^{'})$,
$(n,n^{'})$; some of these observables have been discussed in this work.
Problem of isospin symmetry breaking due to the coupling to the continuum can
be addressed by comparing electromagnetic processes, {\it e.g.},
$B(\Pi \lambda )$ transition matrix elements for certain states in mirror
nuclei, and weak interaction processes like the first-forbidden
$\beta$-decay  in mirror reactions. Finally, for nuclei close and beyond
the proton (neutron) drip lines, the spontaneous proton (neutron)
radioactivity can be studied in the microscopic framework of SMEC (SM).
These unifying features of SMEC approach are extremely useful for
understanding of the structure of exotic nuclei far from the $\beta$ - stability
for which the available experimental information will be scarce.

In this work we have studied nuclei close to the doubly
magic $^{16}\mbox{O}$  in order to understand certain
basic features of the SMEC and, in particular, of the $Q$ - $P$ coupling
operator acting in the restricted SM configuration space. The resulting
quenching of operators ${\cal O}$ and  ${\cal R}$ ( Eqs.  (\ref{ef1}) and 
(\ref{ef2}) respectively)
could be related to the spectroscopic amplitudes for positive parity states in
$^{17}\mbox{F}$ ($^{17}\mbox{O}$) and to the amount of $2p - 2h, 4p - 4h$
correlations in the g.s.\ of $^{16}\mbox{O}$. It was found also that
this SM motivated correction of the effective operator does not solve the
problem of 'halo' of discrete states for positive energies. This problem
results from non-locality of the effective SMEC Hamiltonian and, more
precisely, from the non-hermitean corrections to the eigenvalues for positive
energies which generate the imaginary part. The imaginary part of
eigenvalues and, hence, the size of this halo effect,
is particularly large for pure single-particle
(single-hole) configurations. For this reason, the simplification of structure
of the many body states by neglecting the configuration mixing can in certain
cases lead to  an unphysical enhancement
of resonant-like correction from bound states in,
{\it e.g.}, the elastic cross-section or the elastic phase-shifts.
In this sense, $^{17}\mbox{F}$ and $^{17}\mbox{O}$ nuclei,
with predominantly  s.p.\ structure of positive parity states and extreme
sensitivity to higher order correlations in the many-body wave functions
are somewhat pathological. We believe that this problem will disappear in
nuclei having more particles in the open shells, in which case SM will produce
sufficient amount of mixing in the wave functions.

More complicated decay
channels involving, {\it e.g.}, $\alpha$ particle,
$^3\mbox{He}$ or $^3\mbox{H}$ in the continuum, are beyond
the scope of SMEC in its present form. The future extension of the SMEC for
such
cluster configurations is possible in a framework proposed by Balashov et al
\cite{balashov}. It is encouraging, however, that these possible
shortcomings in the description of decay channels, are so unambiguously
reflected in the calculated decay width for these states \cite{bnop2}.
In general, the decay width is particularly sensitive to the details of the
SM wave functions involved and to the values of matrix elements
of residual coupling so they provide a sensible test of the quality
of SMEC wave functions and/or approximations involved.

The present studies have shown that SMEC results depend sensitively on very
small number of parameters. Some of them, like the parametrization of the
residual interaction which couples states in $Q$ and $P$ subspaces, has been
established in the present work for $sd$-shell nuclei. The others, related
to the quenching of the effective coupling operator can be explained
consistently with the SM analysis of spectroscopic amplitudes. Finally, the
energies of s.p.\ states, 
which determine the radial wave function of many-body states, are
bound by the SM spectroscopic factors and experimental binding energy in
studied nuclei. This gives us a confidence that the 
SMEC can have large predictive
power when applied to the nuclei in the less known regions of the mass table.
The calculations can be performed on a similar level of sophistication as the
SM which, with the recent progress in SM techniques and the effective
interactions has been applied to the medium-heavy nuclei \cite{madstra}.

\vskip 1truecm

{\bf Acknowledgments}\\
We thank E. Caurier for his help in the early stage of development of SMEC
model, and P. Descouvement and R. Morlock for helpful informations.
This work was partly supported by
KBN Grant No. 2 P03B 097 16 and the Grant No. 76044
of the French - Polish Cooperation.


\vfill
\newpage

\input epsf
\begin{table}[h]
\caption{The comparison of amplitudes of
SM  wave functions in five different models for $0_1^+$ state
in $^{16}\mbox{O}$. The entry WBT92 corresponds to the results of Warburton,
Brown and Millener using the '${\Delta}_{4{\hbar}{\omega}}$'--method
\protect\cite{wbm}. The entry WBP($4p4h$) corresponds to the results of SM
calculations using WBP interaction \protect\cite{warburton} and the
'${\Delta}_{4p4h}$'--method \protect\cite{wbm}. Results of the Brown-Green
(BG) \protect\cite{brown}, Zuker-Buck-McGrory (ZBM) \protect\cite{zbm} and
Haxton-Johnson (HJ) \protect\cite{haxton} are also given. }
\label{decompo1}
\begin{center}
\begin{tabular}{|c|c|c|c|}
\hline
 Model & $(0p - 0h)^{J^{\pi}=0_{1}^+}$  & $(2p - 2h)^{J^{\pi}=0_{1}^+}$
& $(4p - 4h)^{J^{\pi}=0_{1}^+}$ \\
\hline
 BG \cite{brown} & 0.874  & 0.469 & 0.130 \\
 ZBM \cite{zbm} & 0.71 & 0.58 & ...\\
 HJ \cite{haxton} & 0.648 & 0.67 & 0.14\\
 WBT92 \cite{wbm} & 0.748  & 0.574 & 0.333 \\
 WBP($4p4h$) \cite{wbm} & 0.775  & 0.557 & 0.299 \\
\hline
\end{tabular}
\end{center}
\end{table}

\begin{table}[h]
\caption{The amplitudes of dominant $(1p - 0h)$ component in the SM wave
function for ${5/2}_{1}^{+}$ (the ground state) and excited states
${1/2}_{1}^{+}$, ${3/2}_{1}^{+}$, calculated by Brown-Green (BG)
\protect\cite{brown} and Zuker-Buck-McGrory \protect\cite{zbm}.}
\label{decompo2}
\begin{center}
\begin{tabular}{|c|c|c|c|}
\hline
 Model & $(1p - 0h)^{J^{\pi}={1/2}_{1}^+}$  & $(1p - 0h)^{J^{\pi}={3/2}_{1}^+}$
& $(1p - 0h)^{J^{\pi}={5/2}_{1}^+}$ \\
\hline
 BG \cite{brown} & 0.881  & 0.718 & 0.901 \\
 ZBM \cite{zbm} & 0.65 & 0 & 0.69\\
\hline
\end{tabular}
\end{center}
\end{table}

\begin{table}[h]
\caption{Experimental spectroscopic amplitudes for
the ground state ${5/2}_{1}^{+}$ and first excited state
${1/2}_{1}^{+}$ }.
\label{decompo3}
\begin{center}
\begin{tabular}{|c|c|c|c|c|}
\hline
 $J^{\pi}$ & Ref. \cite{rolfs} & Ref. \cite{yasue} & Ref. \cite{fortune} &
Ref. \cite{vernotte} \\
\hline
 $5/2^{+}$ & 0.949  & 1.14  & 0.964  & 1  \\
 $1/2^{+}$ & 1  & 0.866  & 0.916  & 0.905  \\
\hline
\end{tabular}
\end{center}
\end{table}

\begin{table}[h]

\caption{The parameters of initial potentials $U(r)$ (\protect\ref{pot})
used in the calculations of self-consistent potentials $U^{(sc)}(r)$ for
the WB  and DDSM1 residual interactions. $U^{(sc)}(r)$ are constructed for
various positive and negative parity states in $^{17}\mbox{F}$ and
$^{17}\mbox{O}$. For all considered cases
the radius of the potential is $R_0=3.214\,$fm and the diffuseness parameter is
$a=0.58\,$fm. The spin-orbit parameter is
$V_{SO}=-3.683\,$MeV. For more details, see the description in the text.}
\label{parameters}
\begin{center}
\begin{tabular}{|c|l|l|l|l|}
\hline
System & $J^{\pi}$ [MeV] & $l j$ & $V_0$ [MeV] (WB) & $V_0$ [MeV] (DDSM1)\\
\hline
[p $\bigotimes$ $^{16}$O]  & $ 5/2^{+} $ & $ 0d_{5/2} $ & $ -42.228 $ & $ -42.123$ \\
                &  $ 1/2^{+} $ & $ 1s_{1/2} $ & $ -44.485 $ & $-45.875$ \\
                &  $ 3/2^{+} $ & $ 0d_{3/2} $ & $ -42.416 $ & $-42.237$ \\
                &  $ 1/2^{-} $ & $ 1p_{1/2} $ & $ -39.448 $ & $-42.330$ \\
                &  $ 3/2^{-} $ & $ 1p_{3/2} $ & $ -39.914 $ & $-42.200$ \\
                &  $ 5/2^{-} $ & $ 0f_{5/2} $ &           &         \\
                &  $ 7/2^{-} $ & $ 0f_{7/2} $ &         &         \\
\hline
[n $\bigotimes$ $^{16}$O]  & $ 5/2^{+} $ & $ 0d_{5/2} $ & $ -42.060 $ & $-42.123$ \\
                &  $ 1/2^{+} $ & $ 1s_{1/2} $ & $ -44.588 $ & $-45.875$ \\
                &  $ 3/2^{+} $ & $ 0d_{3/2} $ & $ -42.440 $ & $-42.237$ \\
                &  $ 1/2^{-} $ & $ 1p_{1/2} $ & $ -41.310 $ & $-42.330$ \\
                &  $ 3/2^{-} $ & $ 1p_{3/2} $ & $ -41.613 $ & $-42.200$ \\
                &  $ 5/2^{-} $ & $ 0f_{5/2} $ & $ -43.921 $ & $-42.213$ \\
                &  $ 7/2^{-} $ & $ 0f_{7/2} $ & $ -43.878 $ & $-42.701$ \\
\hline
\end{tabular}
\end{center}
\end{table}

\begin{table}[h]

\caption{The energies of s.p.\ orbits in the equivalent average
potential $U^{(eq)}(r)$ for protons and neutrons in $^{17}\mbox{F}$
which yield
the same s.p.\ energies as the self-consistent
potentials $U^{(sc)}(r;J^{\pi})$. }
\label{spen}
\begin{center}
\begin{tabular}{|c|l|l|}
\hline
$l j$ & $V_0$ (Protons) & $V_0$ (Neutrons)\\
\hline
$ 0s_{1/2} $ & $-28.168$ & $\cdots $ \\
$ 0p_{3/2} $ & $-14.287$ & $-18.452$ \\
$ 0p_{1/2} $ & $-12.131$ & $-16.297$ \\
$ 0d_{5/2} $ & $-0.6$ & $\cdots $ \\
$ 1s_{1/2} $ & $-0.105$ & $\cdots $ \\
$ 0d_{3/2} $ & $2.497$ & $\cdots $ \\
$ 0f_{7/2} $ & $12.27$ & $\cdots $ \\
$ 0f_{5/2} $ & $22.287$ & $\cdots $ \\
\hline
\end{tabular}
\end{center}
\end{table}

\begin{table}[h]

\caption{The energies of s.p.\ orbits in the equivalent average
potential $U^{(eq)}(r)$ for neutrons and protons in $^{17}\mbox{O}$
which yield the same s.p.\ energies as the self-consistent
potentials $U^{(sc)}(r;J^{\pi})$. }
\label{spen1}
\begin{center}
\begin{tabular}{|c|l|l|}
\hline
$l j$ & $V_0$ (Neutrons) & $V_0$ (Protons)\\
\hline
$ 0s_{1/2} $ & $-32.823$ & $\cdots $ \\
$ 0p_{3/2} $ & $-18.447$ & $-14.8$ \\
$ 0p_{1/2} $ & $-16.378$ & $-12.73$ \\
$ 0d_{5/2} $ & $-4.143$ & $\cdots $ \\
$ 1s_{1/2} $ & $-3.273$ & $\cdots $ \\
$ 0d_{3/2} $ & $-0.91$ & $\cdots $ \\
$ 0f_{7/2} $ & $8.701$ & $\cdots $ \\
$ 0f_{5/2} $ & $16.397$ & $\cdots $ \\
\hline
\end{tabular}
\end{center}
\end{table}

\begin{table}[h]
\caption{SM  energies and SMEC
energies and widths vs.\ experimental ones of $^{17}$F nucleus. 
The proton separation energy is adjusted in order to reproduce
the binding energy of the first excited $1/2^{+}$ state.  
Different labels denote as follows:
'(WB)' -- results of SMEC calculations for the WB interaction (\ref{force})
, '(DDSM1)' -- results of SMEC calculations for the density dependent DDSM1
interaction (\ref{force1}). In these two cases, the quenching factors
(\ref{ef4a} - \ref{ef4c}) for the
positive parity states have been included. Label (DDSM1$^*$) denotes results of
SMEC calculations without the quenching factors. 
Only the g.s. of $^{16}\mbox{O}$ is included in the coupling matrix elements. 
The cut-off radius is $R_{cut} = 9.5\,$fm for the $d_{3/2}$ s.p.\ wave
function in $3/2_{1}^+$ many body state which is in the continuum. 
For the details of the residual interaction which couples $Q$ and $P$ 
subspaces, see the discussion in the text.}
\label{f17}
\begin{center}
\begin{tabular}{|c|c|c|c|c|c|c|c|c|c|}
\hline
$J^\pi$ & $E_{SM}$  & $E$ (WB) & $\Gamma$ (WB) 
& $E$ (DDSM1) & $\Gamma$ (DDSM1) 
& $E$ (DDSM1$^*$) & $\Gamma$ (DDSM1$^*$) 
        & $E_{exp}$  & $\Gamma_{exp}$ \\ 
        &    (MeV)      &  (MeV)    &   (keV)
        &    (MeV)      &   (keV)    
        &    (MeV)      &   (keV)    
        &   (MeV)       &    (keV)            \\
\hline
 \frp5 & $-0.826$ & $-1.467$ & $\cdots $   & $-1.377$ & $\cdots $ & $-1.675$ & $\cdots $ & $-0.600$ & $\cdots $ \\
 \frp1 & -0.105 & $-0.105$ & $\cdots $   & $-0.105$ & $\cdots $ & $-0.105$ & $\cdots $ &$-0.105$  & $\cdots $\\
 \frm1 & 2.134 & 2.486  & 15   & 2.286  & 4.4 & 2.388 & 6.2 & 2.504 & 19 \\
 \frm5 & 3.279 & 3.654  & 0.04 & 3.44  & $\sim$0 & 3.544 & $\sim$0 & 3.257 & 1.5 \\
 \frp3 & 5.044 & 4.085  & 859  & 4.096  & 926  & 2.876 & 848 & 4.400 & 1530 \\
 \frm9 & 3.786 & 4.169  & $\sim 0$ & 3.946  & $\sim 0$ & 4.050 & $\sim 0$ & 4.620 & $\cdots $ \\
 \frm3 & 4.459 & 4.747  & 152  & 4.551  & 120  & 4.636 & 154 & 4.040 & 225 \\
 \frm7 & 5.045 & 5.299  & 9   & 5.172  & 1.2   & 5.256 & 2.2 & 5.072 & 40 \\
 \frm5 & 5.016 & 5.307  & 1  & 5.175  & 0.02  & 5.279 & 0 & 5.082 & $<0.6$ \\
 \frm1 & 5.156 & 5.538  & 2.9  & 5.313  & 5    & 5.416 & 7 & 5.437 & 30 \\
 \frm3 & 5.277 & 5.661  & 2.9  & 5.432  & 14   & 5.535 & 17 & 4.888 & 48 \\
\hline
\end{tabular}
\end{center}
\end{table}

\begin{table}[h]
\caption{The same as in Table~\protect\ref{f17} but for $^{17}$O.
For more details see the description in the text.}
\label{o17}
\begin{center}
\begin{tabular}{|c|c|c|c|c|c|c|c|}
\hline
$J^\pi$ & $E_{SM}$  & $E$ (WB) & $\Gamma$ (WB)
        & $E$ (DDSM1)& $\Gamma$ (DDSM1)
        & $E_{exp}$  & $\Gamma_{exp}$ \\
        &    (MeV)      &   (MeV)    &   (keV)
        &    (MeV)      &   (keV)    &   (MeV)
        &    (keV)            \\
\hline
 \frp5 & $-4.406$ & $-5.245$ & $\cdots $ & $-4.938$ & $\cdots $ & $-4.143$ & $\cdots $\\
 \frp1 & $-3.685$ & $-3.855$ & $\cdots $ & $-3.856$  & $\cdots $ & $-3.273$ & $\cdots $\\
 \frm1 & $-1.446$ & $-1.464$ & $\cdots $ & $-1.450$ & $\cdots $  & $-1.088$ & $\cdots $ \\
 \frm5 & $-0.301$ & $-0.301$ & $\cdots $ & $-0.301$  & $\cdots $ & $-0.301$ & $\cdots $ \\
 \frp3 & 1.464 &0.309  & 83 & 0.543 & 208 &0.942 & 96 \\
 \frm9 & 0.206 & 0.207 & $\sim 0$ & 0.206 & $\sim 0$  & 1.073 & $<0.1$\\
 \frm3 & 0.879 & 0.767  & 49 & 0.804 & 40.4 & 0.410 & 40 \\
 \frm7 & 1.465 & 1.440 & 0.08  & 1.297 & 0.4 & 1.554 & 3.4 \\
 \frm5 & 1.436 & 1.428 & $\sim 0$ & 1.396 & 0.06 & 1.589 & $<1$ \\
 \frm1 & 1.576 & 1.575 & 1.9 & 1.572 & 3.8  & 1.796 & 32 \\
 \frm3 & 1.697 & 1.699 & 1.5 & 1.691 & 8.8  & 1.236 & 28 \\
\hline
\end{tabular}
\end{center}
\end{table}

\begin{table}[h]

\caption{$E1$, $M1$, $E2$ contributions in the limit $E_{CM} \rightarrow 0$
from different incoming waves to the
$S$-factor for the branch
$^{16}\mbox{O}(p,\gamma)^{17}\mbox{F}(J^{\pi}=1/2_1^{+})$. The values
are extracted by fitting second
order polynomial to the calculated SMEC values for the DDSM1 residual coupling.
For more details see the description in the text.}
\label{du1}
\begin{center}
\begin{tabular}{|c|c|c|c|c|c|}
\hline
  \multicolumn{6}{| c |}{$^{16}$O$(p,\,\gamma)^{17}$F${(\frac{1}{2}}^+)$} \\
\hline
\hline
 $\cal ML$ & \multicolumn{2}{ c |}{ E1 } & M1
  & \multicolumn{2}{ c |}{ E2 } \\
\hline
 $\ell_j$ & p$_{1/2}$ & p$_{3/2}$ & s$_{1/2}$ & d$_{3/2}$ & d$_{5/2}$ \\
\hline
 $S(0)$~[MeV$\cdot$b] & $3.06\times 10^{-3}$ & $6.12\times 10^{-3}$
  & $6.93\times 10^{-5}$   & $5.40\times 10^{-7}$ & $8.11\times 10^{-7}$  \\
 $S'(0)/S(0)$~[MeV$^{-1}$] & $-4.95$ & $-4.95$ & $-9.45$ & $0.023$ & $0.025$ \\
\hline
\end{tabular}
\end{center}
\end{table}

\begin{table}[h]

\caption{The same as Table~\protect\ref{du1} for the transitions in the branch
$^{16}\mbox{O}(p,\gamma)^{17}\mbox{F}(J^{\pi}=5/2_1^{+})$.}
\label{du2}
\begin{center}
\begin{tabular}{|c|c|c|c|c|c|}
\hline
  \multicolumn{6}{| c |}{$^{16}$O$(p,\,\gamma)^{17}$F${(\frac{5}{2}}^+)$} \\
\hline
\hline
 $\cal ML$ & \multicolumn{3}{ c |}{ E1 } & \multicolumn{2}{ c |}{ M1 } \\
\hline
 $\ell_j$ & p$_{3/2}$ & f$_{5/2}$ & f$_{7/2}$ & d$_{3/2}$ & d$_{5/2}$ \\
\hline
 $S(0)$~[MeV$\cdot$b] & $2.83\times 10^{-4}$ & $1.11\times 10^{-7}$
   & $2.22\times 10^{-6}$ & $1.43\times 10^{-8}$ & $8.65\times 10^{-7}$ \\
 $S'(0)/S(0)$~[MeV$^{-1}$] & $0.534$ & $7.85$ & $7.86$ & $0.36$ & $3.36$ \\
\hline
\hline
 $\cal ML$ & \multicolumn{5}{ c |}{ E2 } \\
\hline
 $\ell_j$  & s$_{1/2}$ & d$_{3/2}$ & d$_{5/2}$ & g$_{7/2}$ & g$_{9/2}$ \\
\hline
 $S(0)$~[MeV$\cdot$b]  & $3.46\times 10^{-6}$ & $7.60\times 10^{-9}$
  & $3.26\times 10^{-8}$ & $1.6\times 10^{-12}$ & $2.0\times 10^{-11}$ \\
 $S'(0)/S(0)$~[MeV$^{-1}$] & $-17.76$ & $3.15$ & $3.28$ & $18.3$ & $18.3$ \\
\hline
\end{tabular}
\end{center}
\end{table}


\vfill
\newpage

\newpage
\newpage
\newpage
\newpage
\newpage
\newpage
\newpage
\newpage
\newpage
\newpage

\end{document}